\PassOptionsToPackage{unicode}{hyperref}
\PassOptionsToPackage{hyphens}{url}
\PassOptionsToPackage{dvipsnames,svgnames,x11names}{xcolor}
\documentclass[
]{article}
\usepackage{xcolor}
\usepackage{amsmath,amssymb}
\usepackage{iftex}
\ifPDFTeX
  \usepackage[T1]{fontenc}
  \usepackage[utf8]{inputenc}
  \usepackage{textcomp} 
\else 
  \usepackage{unicode-math} 
  \defaultfontfeatures{Scale=MatchLowercase}
  \defaultfontfeatures[\rmfamily]{Ligatures=TeX,Scale=1}
\fi
\usepackage{lmodern}
\ifPDFTeX\else
\fi
\IfFileExists{upquote.sty}{\usepackage{upquote}}{}
\IfFileExists{microtype.sty}{
  \usepackage[]{microtype}
  \UseMicrotypeSet[protrusion]{basicmath} 
}{}
\makeatletter
\@ifundefined{KOMAClassName}{
  \IfFileExists{parskip.sty}{%
    \usepackage{parskip}
  }{
    \setlength{\parindent}{0pt}
    \setlength{\parskip}{6pt plus 2pt minus 1pt}}
}{
  \KOMAoptions{parskip=half}}
\makeatother
\makeatletter
\ifx\paragraph\undefined\else
  \let\oldparagraph\paragraph
  \renewcommand{\paragraph}{
    \@ifstar
      \xxxParagraphStar
      \xxxParagraphNoStar
  }
  \newcommand{\xxxParagraphStar}[1]{\oldparagraph*{#1}\mbox{}}
  \newcommand{\xxxParagraphNoStar}[1]{\oldparagraph{#1}\mbox{}}
\fi
\ifx\subparagraph\undefined\else
  \let\oldsubparagraph\subparagraph
  \renewcommand{\subparagraph}{
    \@ifstar
      \xxxSubParagraphStar
      \xxxSubParagraphNoStar
  }
  \newcommand{\xxxSubParagraphStar}[1]{\oldsubparagraph*{#1}\mbox{}}
  \newcommand{\xxxSubParagraphNoStar}[1]{\oldsubparagraph{#1}\mbox{}}
\fi
\makeatother

\usepackage{color}
\usepackage{fancyvrb}

\DefineVerbatimEnvironment{Highlighting}{Verbatim}{commandchars=\\\{\}}
\usepackage{framed}
\definecolor{shadecolor}{RGB}{241,243,245}
\newenvironment{Shaded}{\begin{snugshade}}{\end{snugshade}}

\newcommand{\AttributeTok}[1]{\textcolor[rgb]{0.40,0.45,0.13}{#1}}

\newcommand{\CommentTok}[1]{\textcolor[rgb]{0.37,0.37,0.37}{#1}}

\newcommand{\ControlFlowTok}[1]{\textcolor[rgb]{0.00,0.23,0.31}{\textbf{#1}}}

\newcommand{\DecValTok}[1]{\textcolor[rgb]{0.68,0.00,0.00}{#1}}
\newcommand{\DocumentationTok}[1]{\textcolor[rgb]{0.37,0.37,0.37}{\textit{#1}}}

\newcommand{\FunctionTok}[1]{\textcolor[rgb]{0.28,0.35,0.67}{#1}}

\newcommand{\KeywordTok}[1]{\textcolor[rgb]{0.00,0.23,0.31}{\textbf{#1}}}
\newcommand{\NormalTok}[1]{\textcolor[rgb]{0.00,0.23,0.31}{#1}}
\newcommand{\OperatorTok}[1]{\textcolor[rgb]{0.37,0.37,0.37}{#1}}
\newcommand{\OtherTok}[1]{\textcolor[rgb]{0.00,0.23,0.31}{#1}}

\newcommand{\SpecialCharTok}[1]{\textcolor[rgb]{0.37,0.37,0.37}{#1}}

\newcommand{\StringTok}[1]{\textcolor[rgb]{0.13,0.47,0.30}{#1}}

\usepackage{longtable,booktabs,array}
\usepackage{calc} 
\usepackage{etoolbox}
\makeatletter
\patchcmd\longtable{\par}{\if@noskipsec\mbox{}\fi\par}{}{}
\makeatother
\IfFileExists{footnotehyper.sty}{\usepackage{footnotehyper}}{\usepackage{footnote}}
\makesavenoteenv{longtable}
\usepackage{graphicx}
\makeatletter
\newsavebox\pandoc@box
\newcommand*\pandocbounded[1]{
  \sbox\pandoc@box{#1}%
  \Gscale@div\@tempa{\textheight}{\dimexpr\ht\pandoc@box+\dp\pandoc@box\relax}%
  \Gscale@div\@tempb{\linewidth}{\wd\pandoc@box}%
  \ifdim\@tempb\p@<\@tempa\p@\let\@tempa\@tempb\fi
  \ifdim\@tempa\p@<\p@\scalebox{\@tempa}{\usebox\pandoc@box}%
  \else\usebox{\pandoc@box}%
  \fi%
}
\def\fps@figure{htbp}
\makeatother

\NewDocumentCommand\citeproctext{}{}

\makeatletter
 \let\@cite@ofmt\@firstofone
 \def\@biblabel#1{}
 \def\@cite#1#2{{#1\if@tempswa , #2\fi}}
\makeatother
\newlength{\cslhangindent}
\newlength{\csllabelwidth}
\newenvironment{CSLReferences}[2] 
 {\begin{list}{}{%
  \setlength{\itemindent}{0pt}
  \setlength{\leftmargin}{0pt}
  \setlength{\parsep}{0pt}
  \ifodd #1
   \setlength{\leftmargin}{\cslhangindent}
   \setlength{\itemindent}{-1\cslhangindent}
  \fi
  \setlength{\itemsep}{#2\baselineskip}}}
 {\end{list}}
\usepackage{calc}

\providecommand{\tightlist}{%
  \setlength{\itemsep}{0pt}\setlength{\parskip}{0pt}}

\usepackage{booktabs}
\usepackage{longtable}
\usepackage{array}
\usepackage{multirow}
\usepackage{wrapfig}
\usepackage{float}
\usepackage{colortbl}
\usepackage{pdflscape}
\usepackage{tabu}
\usepackage{threeparttable}
\usepackage{threeparttablex}
\usepackage[normalem]{ulem}
\usepackage{makecell}
\usepackage{xcolor}
\usepackage[ruled]{algorithm2e}
\usepackage{mathrsfs}
\usepackage{etoolbox}
\AtBeginDocument{\floatplacement{codelisting}{tbp}}
\usepackage{arxiv}
\usepackage{orcidlink}
\usepackage{amsmath}
\usepackage[T1]{fontenc}
\makeatletter
\@ifpackageloaded{caption}{}{\usepackage{caption}}
\AtBeginDocument{%
\ifdefined\contentsname
  \renewcommand*\contentsname{Table of contents}
\else
  \newcommand\contentsname{Table of contents}
\fi
\ifdefined\listfigurename
  \renewcommand*\listfigurename{List of Figures}
\else
  \newcommand\listfigurename{List of Figures}
\fi
\ifdefined\listtablename
  \renewcommand*\listtablename{List of Tables}
\else
  \newcommand\listtablename{List of Tables}
\fi
\ifdefined\figurename
  \renewcommand*\figurename{Figure}
\else
  \newcommand\figurename{Figure}
\fi
\ifdefined\tablename
  \renewcommand*\tablename{Table}
\else
  \newcommand\tablename{Table}
\fi
}
\@ifpackageloaded{float}{}{\usepackage{float}}
\floatstyle{ruled}
\@ifundefined{c@chapter}{\newfloat{codelisting}{h}{lop}}{\newfloat{codelisting}{h}{lop}[chapter]}
\floatname{codelisting}{Listing}

\usepackage{amsthm}
\theoremstyle{plain}
\newtheorem{proposition}{Proposition}[section]
\theoremstyle{definition}
\newtheorem{definition}{Definition}[section]
\theoremstyle{remark}
\AtBeginDocument{}

\makeatother
\makeatletter
\@ifpackageloaded{caption}{}{\usepackage{caption}}
\@ifpackageloaded{subcaption}{}{\usepackage{subcaption}}
\makeatother
\usepackage{bookmark}
\IfFileExists{xurl.sty}{\usepackage{xurl}}{} 
\hypersetup{
  pdftitle={Crossmaps: standardizing ex-post data harmonization workflows},
  pdfauthor={Cynthia A. Huang},
  colorlinks=true,
  linkcolor={blue},
  filecolor={Maroon},
  citecolor={Blue},
  urlcolor={Blue},
  pdfcreator={LaTeX via pandoc}}

\newcommand{\runninghead}{A Preprint }
\title{Crossmaps: standardizing ex-post data harmonization workflows}
\author{\textbf{Cynthia A. Huang}~\orcidlink{0000-0002-9218-987X}\\\\LMU
Munich, and Munich Center for Machine Learning
(MCML)\\\\\\\\\\\href{mailto:cynthia.huang@lmu.de}{cynthia.huang@lmu.de}}
\date{}
\begin{document}
\maketitle
\begin{abstract}
Ex-post harmonization, whereby data collected under one classification
standard are reclassified and redistributed under another to facilitate
joint analysis is often treated as a simple data preparation task, but
it is in fact complex imputation. When a value is split across several
target categories, the harmonized values rest on assumptions that should
be stated, checked, and carried through to downstream analysis.
Unfortunately, these assumptions are often hidden in custom data
wrangling scripts and seldom systematically checked. The Crossmaps
framework separates the specification and implementation of
transformation logic into two new data structures based on a new task
abstraction for mapping aggregate statistics from one classification to
another. The part-to-whole array holds a total together with its
distribution across a set of keys, and the crossmap holds the
redistribution logic. Combined they define the block-level operation of
a crossmap transform, which imputes values reported in a source
classification into a target classification. We give equivalent graph,
matrix and edge list encodings, each supporting different inspection,
validation and extraction tasks, to improve the transparency and reuse
of harmonization efforts. We include a demonstration based a published
ex-post harmonised industrial-statistics dataset using the \texttt{xmap}
package.
\end{abstract}

\section{Introduction}\label{sec-introduction}

Statistical software represents a great deal about data. A variable can
be declared a factor with known levels, a date can be typed as a date,
and a value can be recorded as absent rather than zero. However,
comparatively little is represented about how statistics are mapped from
one classification standard to another: which source categories were
split, in what proportions, and on whose judgment. This is a
consequential omission, because harmonized datasets are not simply
cleaned and integrated versions of existing datasets. Ex-post
harmonization, whereby existing data collected under one classification
standard are reclassified and redistributed under another classification
to facilitate joint analysis, is in fact complex imputation. In the case
of hierarchical classifications that aggregate up to some notion of a
shared sum (e.g.~occupation counts or industry-level output statistics),
the ``clean'' dataset is actually an estimate of what would have been
collected under a shared target classification, such as an international
rather than national standard classification. As with any estimate, it
rests on assumptions that should be stated, checked, and carried through
to downstream analysis.

Unfortunately, the details of a harmonization strategy are often hidden
away in custom data wrangling scripts, and only described in general
terms as part of the data preparation process. Mapping details are often
relegated to footnotes, appendices or supplementary materials, if
recorded at all. For example, Humlum (2022) harmonizes and integrates
Danish micro-data with occupation codes from the 1988 and 2008 versions
of the Statistics Denmark's Classification of Occupations (DISCO88 and
DISCO08) to study interactions between robot adoption and labor market
dynamics. In the absence of an officially published DISCO88 to DISCO08
correspondence, Humlum combines multiple published correspondences from
both the International Labour Organization and Statistics Denmark with
relations inferred from job code changes in the microdata. Detailed
notes on the DISCO88-DISCO08 correspondence created for and used to
prepare the analysis data are not included in the main paper or
appendix, and can only be found in a separate documentation note on the
author's website (Humlum 2021).

Harmonized datasets reflect subjective decisions that are central to the
epistemological validity of downstream statistical analysis. For
example, when harmonizing occupation counts across time, the choice of
mapping between a new occupation such as ``data scientist'' and
occupations in older surveys (e.g.~``statistician'' or ``computer
programmer''), directly impacts the interpretation and uncertainty
quantification of statements about the growth or decline of statistics
or computer programming. The resultant combined datasets are
``imputations'' of what would have been collected if the same survey
instrument were used in data collection. In the case of harmonizing
occupation lists, the imputation reflects assumptions about how people
who reported themselves as ``data scientists'', would have classified
themselves if ``data scientist'' was available as an option. It follows
that in a harmonized dataset, counts for ``statisticians'' and
``computer programmers'' reflect some proportional reallocation from
``data scientist'' and imputation assumption of the form of statements
like ``50\% of data scientists would have classified themselves as a
`statistician' if given the choice.'' The imputation is, however,
constrained by the fact that the total count across all occupation
categories should be the same before and after any redistribution or
recoding.

Currently, documenting harmonization details and checking that the
imputation is appropriately constrained is relatively ad-hoc. Even when
preparation scripts and reproducible workflows are diligently provided,
the length and complexity of such scripts grows significantly with the
size and number of classifications requiring harmonization. Combined
with the idiosyncrasies of different coding languages and data
manipulation tools, reproducibility alone is not sufficient to
facilitate comprehensive auditing or understanding of an integrated
dataset within a reasonable time frame and amount of effort. Existing
data integration tools and workflows lack formal and standardized
support for specifying, implementing, validating and documenting
harmonization procedures. Although data integration and harmonization
use similar computational data wrangling tools, harmonization requires
support for more than just manipulations of data values. Unified
computational support for harmonization requires maintaining information
about the relationships between the underlying measurement constructs
(i.e.~statistical classifications or other hierarchical structures).

The Crossmaps framework provides this support by separating the
specification of transformation logic from implementation and
manipulation of data using equivalent graph, matrix and relational
algebra representations of mapping statistics from one classification to
another. The Crossmaps framework contributes two declared data
structures for ex-post harmonization --- the \emph{part-to-whole array},
which holds a total together with its distribution across a set of keys,
and the \emph{crossmap}, which holds the redistribution logic ---
together with the conditions under which a transformation can be
validated before it is applied. We base these structures upon a new task
abstraction and formalized a conditions for encoding and validating
\emph{crossmap transforms}, which are transformations of aggregated
statistics from one classification standard to another. Consequently,
properties ordinarily established by inspecting transformed output
become structural conditions on the crossmap itself, and the
transformation persists as an object rather than dissolving into a
script, so the degree to which any target value is imputed rather than
observed remains recoverable.

\section{Related Work}\label{sec-related-work}

Retrospectively reconciling already-collected data is variously called
data fusion, data integration, or data harmonization. However, we use
the term ex-post data harmonization to refer to both the conceptual
reconciliation between related datasets and the computational
manipulation of data. Notably, we separate \emph{data harmonization}
from \emph{data integration}, through the inclusion of conceptual
concerns and computational concerns that \emph{ex-ante harmonization}
can circumvent by designing comparable instruments prior to collection
(Ehling 2003; Granda et al. 2010).

Our work directly answers repeated calls from the field of survey data
harmonization for guidelines and software that standardizes the
documentation of ex-post harmonized procedures and datasets (Granda et
al. 2010; Dubrow and Tomescu-Dubrow 2016; Fortier et al. 2016; Ehling
2003), as well as broader discussions about computational
reproducibility and replicability. However, existing guidelines either
seldom provide integrated conceptual and computational guidance. For
instance, Fortier et al. (2016) discusses survey design considerations
and the need to ensure comparability of measures at length, but
relegates data processing to being ``achieved using algorithms'',
followed by unspecified ad-hoc quality checks and verification of said
algorithms. Similarly, prior attempts to formalize ex-post harmonization
operations as domain-specific language and semantic models (Goerlich and
Ruiz 2018; Denk and Froeschl 2004) lack robust software designs or
implementations. On the other hand, standardized documentation templates
such as data information cards (e.g. Gebru et al. 2021; Pushkarna et al.
2022), and metadata standards (e.g. Koren et al. 2022) are designed for
broad capture of data provenance information rather than the specific
decisions and assumptions relevant to ex-post harmonization.

\begin{figure}

\begin{minipage}[t]{0.57\linewidth}

\begin{longtable}[]{@{}llll@{}}
\caption{Example crosswalk \emph{one-to-one} mapping between the two,
three and numeric country codes from the 2020 release of the
\emph{ISO-3166 International Standard for country codes and codes for
their subdivisions}}\label{tbl-iso-crosswalk}\tabularnewline
\toprule\noalign{}
Country & ISO2 & ISO3 & ISONumeric \\
\midrule\noalign{}
\endfirsthead
\toprule\noalign{}
Country & ISO2 & ISO3 & ISONumeric \\
\midrule\noalign{}
\endhead
\bottomrule\noalign{}
\endlastfoot
Afghanistan & AF & AFG & 004 \\
Albania & AL & ALB & 008 \\
Algeria & DZ & DZA & 012 \\
American Samoa & AS & ASM & 016 \\
Andorra & AD & AND & 020 \\
\end{longtable}

\end{minipage}%
\begin{minipage}[t]{0.05\linewidth}
~\end{minipage}%
\begin{minipage}[t]{0.38\linewidth}

\begin{longtable}[]{@{}llr@{}}
\caption{Hypothetical crossmap for \emph{many-to-many} recoding and
redistribution of national accounts (e.g.~GDP) across years to create
harmonized panel dataset in terms of target states for analysis:
\protect\texttt{BEL}, \protect\texttt{LUX}, \protect\texttt{DEU} and
\protect\texttt{AUS}.}\label{tbl-ctry-crossmap}\tabularnewline
\toprule\noalign{}
from & to & weight \\
\midrule\noalign{}
\endfirsthead
\toprule\noalign{}
from & to & weight \\
\midrule\noalign{}
\endhead
\bottomrule\noalign{}
\endlastfoot
BLX & BEL & 0.5 \\
BLX & LUX & 0.5 \\
E.GER & DEU & 1.0 \\
W.GER & DEU & 1.0 \\
AUS & AUS & 1.0 \\
\end{longtable}

\end{minipage}%

\end{figure}%

Kołczyńska (2022) attempts to address this gap in specific
implementation guidance by adapting existing database schema-matching
practices. They propose the use of annotated lookup tables to encode
concept mappings between two related measures. Such lookup tables
consist of at least two columns, one with keys in the source measure and
one for the target measure, and are variously referred to as
\emph{correspondence} or \emph{concordance} tables in economics and
official statistics (e.g. Pierce and Schott 2012; Dorner and Harhoff
2018), or \emph{Metadata or Schema crosswalks} in database and computing
contexts (Khan et al. 2015; Cheney et al. 2009, 430). As shown in
Table~\ref{tbl-iso-crosswalk} crosswalk tables structure `one-to-one'
recoding logic in a format that is both natural for people to read, and
can also store metadata such as extended descriptions or notes.
Furthermore, as lookup tables, they can be used to transform data
without any additional reshaping or row-wise translation into
programming commands. However, the two-column structure is unable to
support transformations where a single source key is related to multiple
targets, as shown in rows 1 and 2 of Table~\ref{tbl-ctry-crossmap}, and
also known as one-to-many relations. As such, crosswalk-based
computational tools generally treat one-to-many relations as special
cases (Liao et al. 2020; Arel-Bundock et al. 2018; Mackey et al. 2023).
For example, the concordance package (Liao et al. 2020), which maps
between the HS, SITC, NAICS, ISIC and SIC code systems, returns the
first matching target by default, and although it can instead return
shares across all matches, applying those shares is left to the analyst
rather than being part of the transformation. These special cases
reintroduce the need for bespoke code, hindering the auditing and reuse
of the harmonized dataset and increasing the potential for mistakes.

Unfortunately, the existing data manipulation tools used to write such
bespoke code and provenance tracking tools for examining data wrangling
code only partially address the requirements of ex-post harmonization.
In particular, because ex-post harmonization of hierarchical
part-to-whole statistics involves block-level transformations over a
group of rows, existing column-wise or row-wise operations and
constraints (Wickham 2007, 2014) are insufficient to guarantee
group-level properties like the preservation of group totals. Such
constraints can be implemented as assertions or other checks within the
pipeline ({van der Loo and de Jonge} 2021; Fischetti 2023; Iannone et
al. 2026), though what checks are necessary is up to the user to
determine. Moreover, different checks provide different amounts of
diagnostic information -- comparing totals can tell you that something
went wrong, but does not directly locate the problematic source-target
pairings or errors in the code. Consequently, provenance tracking tools
based on comparing tables before and after transformation (e.g.,
Lucchesi et al. 2022; Xiong et al. 2023; Wang et al. 2022; Niederer et
al. 2018) are also of limited utility since differences between input
and output data frames cannot, in general, uniquely identify the mapping
used to produce the output. This is because one-to-many redistributions
can be offset by many-to-one aggregations. Some prior work from the
database research mentions the challenge of ambiguous transformations,
but generally treat the problem as uncommon or impossible ({Kandel et
al.} 2011; Niederer et al. 2018).

The most closely related prior work for the Crossmap framework spans
both official statistics and statistical computing. The closest prior
formalization is Hulliger (1998)'s characterization of classification
linkage as linear maps and correspondence matrices, and accompanying
conditions for verifying linkage. We derive equivalent conditions using
the graph and database representations, and additionally propose
accompanying data structures and operations. The accompanying data
structures and operations most closely resemble prior work in tidy data
structures and graphical grammars (Wickham 2014, 2010; Wang et al.
2020). Similar to Bartonicek et al. (2025)'s identification of validity
conditions for linked selection in interactive data visualization using
algebraic structures, the Crossmaps framework leverages simple
mathematical equivalences to unify the specification, validation and
implementation of ex-post harmonization in a single inspectable object,
so that the conditions a transformation must satisfy can be checked
before it is applied rather than inferred from its output.

\section{The Crossmaps Framework}\label{sec-framework}

\subsection{Ex-Post Harmonization Task
Abstraction}\label{ex-post-harmonization-task-abstraction}

In order to derive validation conditions for ex-post harmonization
operations, we need to define the objects upon which the conditions need
to hold. We construct these objects based on a \emph{provenance task
abstraction} (Bors et al. 2019) of ex-post harmonization focused on
grouping and sequencing conceptually related dataset design and
corresponding implementation tasks, rather than describing common
workflow steps or checklists of required decisions (c.f. Granda and
Blasczyk 2016; Fortier et al. 2016; Kołczyńska 2020).

Our proposed task abstraction comprises four sequential preparation
stages:

\begin{enumerate}
\def\labelenumi{\arabic{enumi}.}
\tightlist
\item
  \textbf{Dataset Conceptualization and Initial Collection:} designing
  the harmonized dataset, and collecting datasets containing relevant
  data.
\item
  \textbf{Source Specific Cleaning:} identifying and resolving issues
  specific to different datasets, data sources and/or collection
  methods.
\item
  \textbf{Crossmap Transforms:} designing and implementing the
  harmonization of each source dataset into a common measurement
  standard.
\item
  \textbf{Data Merging:} merging transformed data into a single
  analysis-ready dataset.
\end{enumerate}

Although source-specific data wrangling and harmonization focused
transformations and merging use similar computational tools, the
conceptual scope and nature of provenance decisions in these two stages
are meaningfully distinct. Source-specific preparation will usually
involve reshaping data without modifying values, as well as row-wise or
column-wise modifications such as imputing missing cell values or
identifying anomalous values or observations. Documentation for this
stage is can generally be organized by source dataset. On the other
hand, \emph{crossmap transforms} involve decisions across multiple
datasets, and thus necessitate provenance objects that can effectively
represent block-wise and cross-table operations and mappings. General
purpose data wrangling tools lack this expressive capacity.

\subsection{Crossmaps and Crossmap
Transforms}\label{crossmaps-and-crossmap-transforms}

\begin{figure}

\centering{

\pandocbounded{\includegraphics[keepaspectratio]{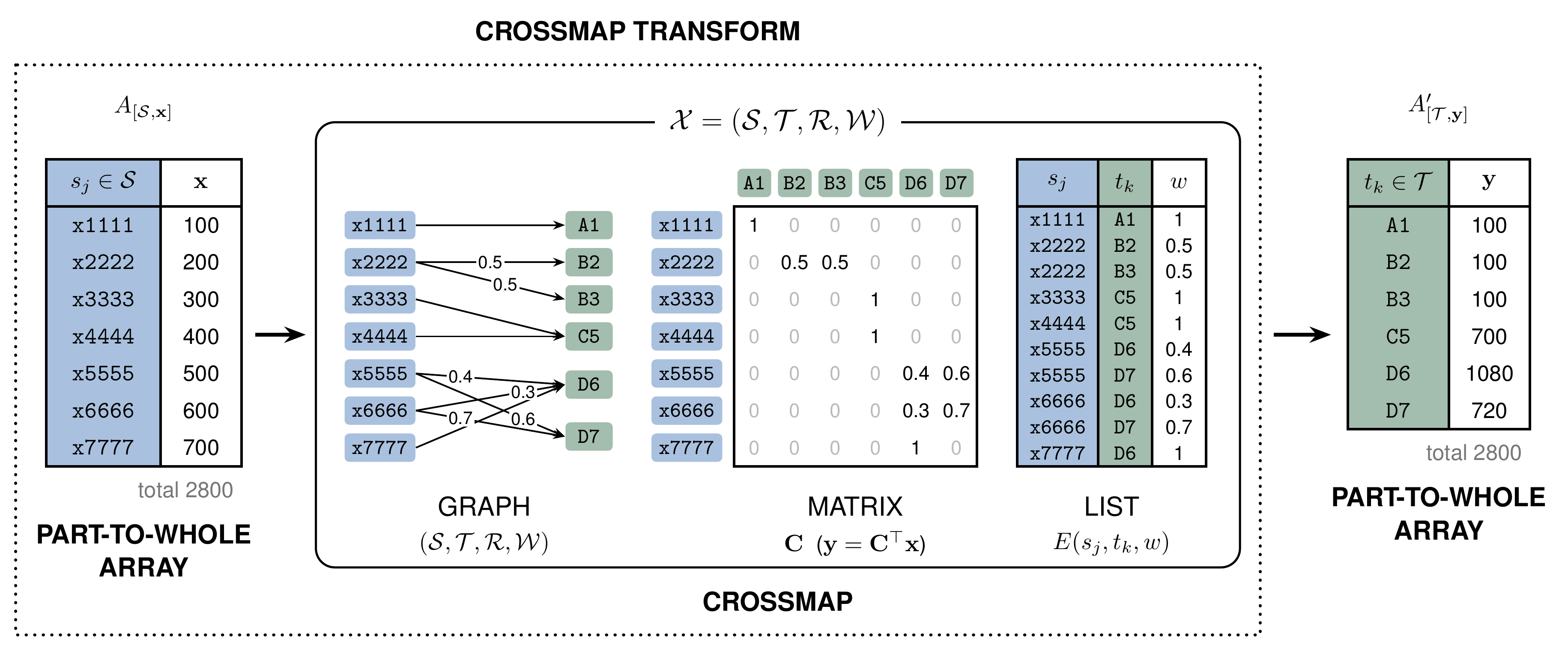}}

}

\caption{\label{fig-crossmap-transform}Conceptual illustration of a
\emph{crossmap transform} (Definition~\ref{def-crossmap-transform}). We
denote the reported part-to-whole observation as
\(A_{[\mathcal{S},\mathbf{x}]}\), and the imputed observation as
\(A'_{[\mathcal{T},\mathbf{y}]}\), where the prime marks values that are
imputed rather than reported. The relationship between \(A\) and \(A'\)
is given by the \emph{crossmap}
\(\mathcal{X} = (\mathcal{S}, \mathcal{T}, \mathcal{R}, \mathcal{W})\)
defined in Definition~\ref{def-crossmap-collection}. The crossmap is
shown in its three equivalent encodings: as a weighted bipartite graph,
as the matrix \(\mathbf{C}\), and as the edge list \(E(s_j,t_k,w)\).}

\end{figure}%

We address the representational gap by defining input and output objects
and conditions for \textbf{\emph{crossmap transforms}}, as shown in
Figure~\ref{fig-crossmap-transform}. A \emph{part-to-whole array} holds
a total together with the distribution of that total across an index,
and a \emph{crossmap} holds the logic by which that total is
redistributed. Neither is mathematically novel, being a non-negative
vector on a finite index and a row-stochastic matrix respectively.
However, each carries a constraint which the underlying type does not
express, and it is these constraints which render coverage of the source
index, the handling of missing values, and the addition or removal of
keys checkable in advance rather than diagnosed after the fact
(Section~\ref{sec-guarantees}).

\begin{definition}[]\protect\hypertarget{def-input-array}{}\label{def-input-array}

A \textbf{part-to-whole array} with index set
\(\mathcal{K} = {\kappa_i : i = 1\dots K}\), is an associative array of
\(K\) key-value pairs, such that \(A_{[\mathcal{K},\mathbf{x}]}\) =
\(\{(\kappa_i, x_i) : \kappa_i \in \mathcal{K}, x_i \in \mathbb{R}_{\geq 0}\}\),
where \(x_i = A(\kappa_i)\) is the non-negative real value retrievable
by the key \(\kappa_i\) and:

\begin{itemize}
\tightlist
\item
  each key \(\kappa_i\) corresponds to part of the conceptual unit
  defined by the index set \(\mathcal{K}\) (e.g.~state in a country),
  and
\item
  the sum \(\sum_{i=1}^{K} x_i\) forms a numeric mass belonging to the
  same unit (e.g.~GDP in each state).
\end{itemize}

\end{definition}

We could also define \(A_{[\mathcal{K},\mathbf{x}]}\) as the function
\(A: \mathcal{K} \to \{x\}\), where \(\{x\}\) is the set of unique
\(x_i\) values. However, defining part-to-whole arrays as associative
arrays of key-value pairs more closely aligns with the tabular format by
which such data are generally presented and shared.

\begin{definition}[]\protect\hypertarget{def-crossmap-collection}{}\label{def-crossmap-collection}

A \textbf{crossmap} is a collection
\(\mathcal{X} = (\mathcal{S}, \mathcal{T}, \mathcal{R}, \mathcal{W})\)
with elements satisfying the following:

\begin{itemize}
\tightlist
\item
  \(\mathcal{S} = \{s_j: j = 1 \dots S\}\) and
  \(\mathcal{T} = \{t_k: k = 1 \dots T\}\) are two sets, referred to as
  the source and target key sets respectively;
\item
  \(\mathcal{R} = \{ (s_j,t_k) : s_j \in \mathcal{S} \text{ shares value with }  t_k \in \mathcal{T} \} \subseteq \mathcal{S} \times \mathcal{T}\)
  is a binary relation between source and target keys, such that there
  exists \((s_j, t_k) \in \mathcal{R}\) for all source keys
  \(s_j \in \mathcal{S}\); and
\item
  \(\mathcal{W} = \{w_{jk} \in (0,1] \text{ if } (s_j,t_k) \in \mathcal{R}: \forall j \ \sum_{k} w_{jk} = 1 \} \subseteq (0,1]^{S \times T}\),
  is a set of weights representing the share of value attached to a
  source key to be distributed to the target key.
\end{itemize}

\end{definition}

Weights will only be fractional in the case of redistribution from a
single source key to multiple target keys, and must total one across all
pairs originating from a given source key.

\begin{definition}[]\protect\hypertarget{def-crossmap-transform}{}\label{def-crossmap-transform}

A \textbf{crossmap transform} is an operation that applies a
\emph{crossmap}
\(\mathcal{X} = (\mathcal{S}, \mathcal{T}, \mathcal{R}, \mathcal{W})\)
to a part-to-whole array \(A_{[\mathcal{K},\mathbf{x}]}\), where
\(\mathcal{K} \subseteq \mathcal{S}\). The operation redistributes the
total numeric mass \(\sum_{i=1}^{K} x_i\) across a target index
\(\mathcal{T}\) and returns a part-to-whole array:
\(A'_{[\mathcal{T},\mathbf{y}]} = \{(t_k, y_k): t_k \in \mathcal{T}, y_k = \sum_{i : (\kappa_i,t_k) \in \mathcal{R}} x_i w_{ik}\}\)

\end{definition}

For an input \(A_{[\mathcal{K},\mathbf{x}]}\) and crossmap
\(\mathcal{X}\) to form a valid crossmap transform, the crossmap must
specify mappings for all of the index keys in the input
(i.e.~\(\mathcal{K} \subseteq \mathcal{S}\)). We refer to this condition
as \textbf{\emph{conformability}}.

Because the weights attached to each source key sum to one, a crossmap
transform preserves the total numeric mass: the sum of the imputed
values \(\sum_{k=1}^{T} y_k\) equals the sum of the reported values
\(\sum_{i=1}^{K} x_i\). We refer to both the condition
\(\forall j \ \sum_{k} w_{jk} = 1\) on the crossmap and the resulting
property of the operation as the \textbf{\emph{mass-preserving
condition}}.

\subsection{Validating ex-post harmonization}\label{sec-guarantees}

Stated over these objects, three properties of crossmap transforms that
are ordinarily established by inspecting transformed output become
conditions on the crossmap itself. \textbf{Sufficient coverage} of the
source part-to-whole array by the crossmap such that
\(\mathcal{K} \subseteq \mathcal{S}\) ensures no values are `left
behind'. \textbf{Missing value arithmetic} errors can be avoided via
conditions on target keys with more than one incoming source value.
\textbf{Addition and removal of index keys} is excluded from crossmap
transforms by construction. Since a crossmap redistributes a fixed total
and cannot create or destroy mass, key-value pairs should only be
removed before a transform and appended after it. Each of these
conditions can be checked with simple comparisons and missing data
detection prior to any actual transformation of data.

\section{Alternative encodings}\label{sec-encodings}

Crossmaps are formally defined in
Definition~\ref{def-crossmap-collection} as collections of sets, each
holding a particular type of information required to specify a mapping
from one statistical classification to another. As shown in
Figure~\ref{fig-crossmap-transform}, there are at least three
alternative representations of the crossmap source keys
(\(\mathcal{S}\)), target keys (\(\mathcal{T}\)), links
(\(\mathcal{R}\)) and weights (\(\mathcal{W}\)). Reading \(\mathcal{S}\)
and \(\mathcal{T}\) as disjoint node sets and \(\mathcal{R}\) as
directed edges labeled by \(\mathcal{W}\) recovers the graph
representation directly from the set form, with no re-encoding required.
The matrix form, which we denote with \(\mathbf{C}\), is a genuine
re-encoding: it fuses \(\mathcal{R}\) and \(\mathcal{W}\) into a single
\(S \times T\) matrix, where links are encoded as the pattern of
non-zero entries and the weights \((0,1]\) as their values, while
\(\mathcal{S}\) and \(\mathcal{T}\) are demoted from keys to row and
column positions. The edge list \(E\) is likewise a re-encoding,
separating the links as rows, with columns for source, target keys and
weights.

Each representation expresses the same information, but the choice of
representation determines how and what a practitioner can easily
inspect, edit and audit. Crossmap conditions and properties can be
intuitive and cheap to check in one representation but awkward in
another. Graphs and node-link visualizations provide an intuitive lens
for understanding and communicating characteristics of crossmap
transforms. The matrix encoding\footnote{Prior independent work shows
  that linkages between statistical classification can be characterized
  as linear mappings between source and target vector spaces (Hulliger
  (1998)).}clarifies the validation conditions for crossmap and
conformability of crossmap transforms. The edge list representation
connects the abstract implications of the framework with existing
statistical software design principles and challenges for tabular data
analysis.

\subsection{Reading Crossmaps as Graphs}\label{sec-graph-encoding}

\begin{figure}

\centering{

\pandocbounded{\includegraphics[keepaspectratio]{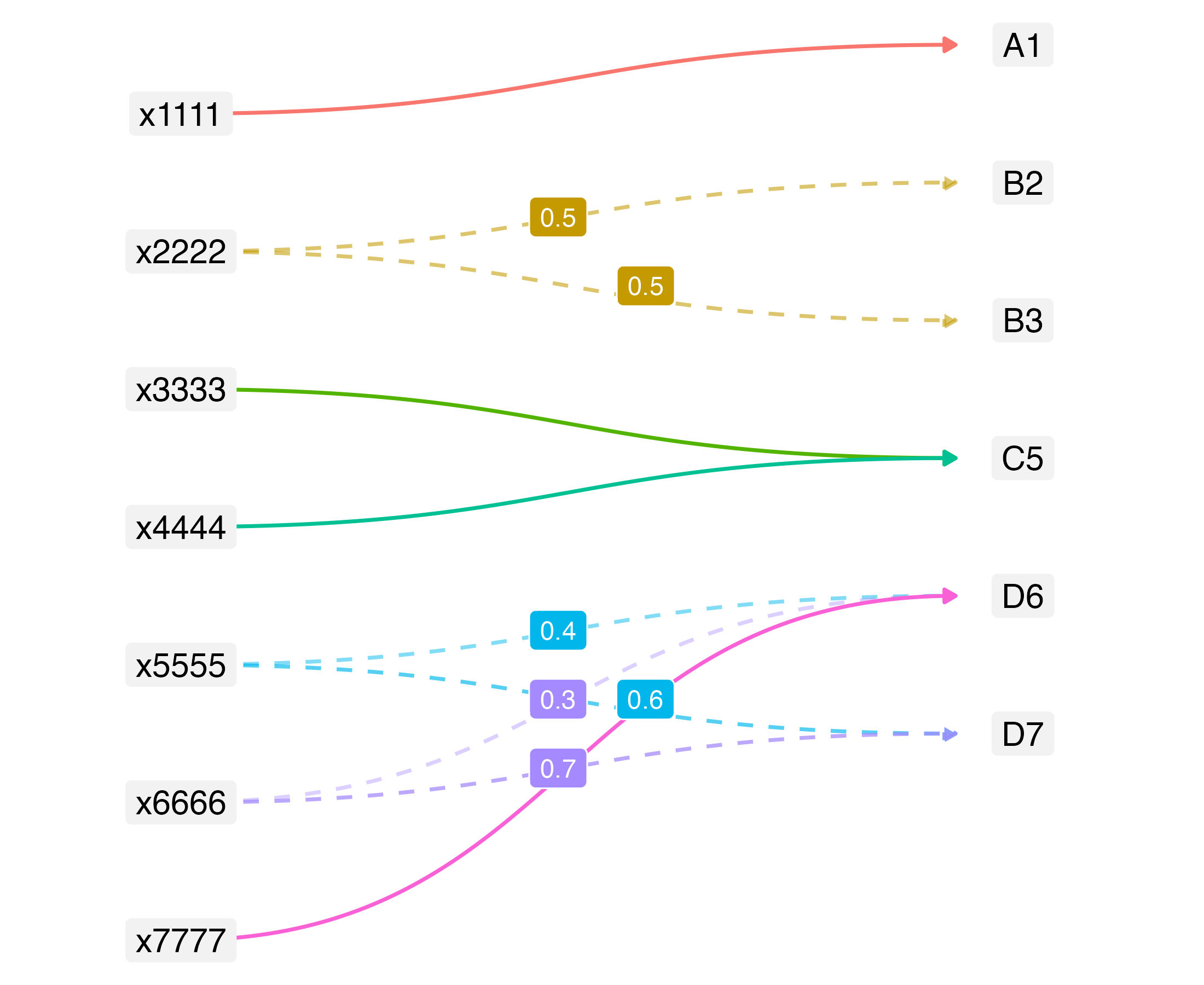}}

}

\caption{\label{fig-simple-bigraph}The top component
\texttt{x1111-\textgreater{}A1} shows a \emph{one-to-one} mapping, the
next subgraph \texttt{x2222-\textgreater{}\{B2,B3\}} is a
\emph{one-to-many} split, the third subgraph
\texttt{\{x3333,x4444\}-\textgreater{}C5} is a \emph{many-to-one}
aggregation. The final subgraph contains overlapping \emph{one-to-one}
and \emph{one-to-many} relationships, composing the ancillary
\emph{many-to-many} relationship type.}

\end{figure}%

Node-link graph visualization are a natural framework for detailed
exploration and editing of crossmap transforms.
Figure~\ref{fig-simple-bigraph} shows a proposed node-link visualization
design for crossmaps highlighting and communicating mapping decisions
(Huang 2023). Line style, ordering, opacity and labels help the view to
focus their attention on links that warrant closer inspection. The
direction of edges directly reflect lateral (one-way) nature of the
mass-preserving condition. Crossmaps constrain the weights leaving each
source key but not what arrives at a target key. Reversing the
aggregation \texttt{\{x3333,\ x4444\}} \(\to\) \texttt{C5} in
Figure~\ref{fig-simple-bigraph} would give \texttt{C5} two outgoing
edges of weight one, so the reversal is not a crossmap. By comparison,
crosswalks are bi-lateral, since they only record the binary relation
\(\mathcal{R}\) between source and target keys.

The four distinct components in Figure~\ref{fig-simple-bigraph}
illustrate the commonly described mapping cases of one-to-one,
one-to-many, many-to-one and many-to-many. The relative complexity of
the \emph{many-to-many} subgraph
\texttt{\{x5555,x6666,x7777\}-\textgreater{}\{D6,D7\}} provides visual
intuition for quantifying and locating consequential imputation
assumptions within a crossmap transform. In particular, the three
incoming links to \texttt{D6} show that transformed value carries
stronger imputation assumptions than the one-to-one transformed value
for \texttt{A1}. Let us define \emph{imputation potential} for a
particular target value in terms of the incoming links such that
\(y_{\mathtt{D6}} = x_{\mathtt{x7777}} + 0.3 x_{\mathtt{x6666}} + 0.4 x_{\mathtt{x5555}}\),
clearly has more imputation potential than
\(y_{\mathtt{A1}} = x_{\mathtt{x1111}}\). The actual degree of
imputation embedded in a harmonized dataset is determined by both the
imputation potential, and the actual input data. For example, Consider
two potential part-to-whole array inputs to the crossmap in
Figure~\ref{fig-simple-bigraph}, one with the majority of the overall
shared total in \texttt{x1111}, and one with the majority in
\texttt{x5555}. In the former case, most of the mass is just re-indexed
as \texttt{A1}, while in the latter case, the value for \texttt{x5555}
is split up between \texttt{\{D6,D7\}}, producing an output array with
much stronger counterfactual assumptions.

The links shown above also illustrate the potential for data loss in
crossmap transforms through missing value arithmetic errors. Missing
values as link weights are precluded by definition. However, errors can
arise if passing a missing value into a one-to-many split, or adding
missing values to non-missing values in a many-to-one relationship.
Imagine that in the input part-to-whole array \(A\), the value for
\texttt{x3333} was missing, but \texttt{x4444} was not. Then value of
\texttt{C5} depends on how the missing value is handled. The missing
value can be destructive if
\(x_{\mathtt{x3333}} + x_{\mathtt{x4444}} = NA\), such that \texttt{C5}
is imputed as missing, and the observed value of \texttt{x3333} is
completely lost. This would also violate the requirement that the total
of \(A\) and \(A'\) match. Alternatively, the value for \texttt{x4444}
could be coerced to zero, or dropped entirely from the input array. Both
are imputation decisions that should ideally be handled explicitly prior
to the crossmap transform for full provenance transparency. On the other
hand, it seems reasonable to propagate a missing value from
\texttt{x1111-\textgreater{}A1}, or
\texttt{x2222-\textgreater{}\{B2,\ B3\}}. However, the incoming links to
\texttt{D6} show that \emph{one-to-one}
(e.g.~\texttt{x7777-\textgreater{}D6}) and \emph{one-to-many}
(e.g.~\texttt{x5555-\textgreater{}D6}) relationships can be combined in
the same target resulting in the same issues as
\texttt{\{x3333,x4444\}-\textgreater{}C5}.

\subsection{Encoding Crossmaps as Matrices}\label{sec-matrix-encoding}

\begin{definition}[]\protect\hypertarget{def-matrix-encoding}{}\label{def-matrix-encoding}

The \textbf{matrix encoding} of
\(\mathcal{X} = (\mathcal{S}, \mathcal{T}, \mathcal{R}, \mathcal{W})\)
is the \(S \times T\) matrix
\(\mathbf{C} = [c_{jk} : c_{jk} = w_{jk} \in \mathcal{W} > 0\text{ if } (s_j,t_k) \in \mathcal{R}, \text{ and } c_{jk} = 0 \text{ otherwise}]\).

\end{definition}

\begin{proposition}[]\protect\hypertarget{prp-transform-linalg}{}\label{prp-transform-linalg}

For a given crossmap transform of the part-to-whole array
\(A_{[\mathcal{K}, \mathbf{x}]}\) (Definition~\ref{def-input-array}) by
a conformable crossmap
\(\mathcal{X} = (\mathcal{S}, \mathcal{T}, \mathcal{R}, \mathcal{W})\)
with matrix encoding \(\mathbf{C}\) resulting in
\(A_{[\mathcal{T}, \mathbf{y}]}\), if \(\mathcal{K}\) and
\(\mathcal{S}\) are identical ordered sets, then
\(\mathbf{y} = \mathbf{C}^{\top}\mathbf{x}\) is equivalent to
\(A_{[\mathcal{T}, \mathbf{y}]}\), where \(\mathbf{x}\) is the vector of
all values \(x_i = A(\kappa_i)\).

\end{proposition}

\(\mathbf{C}\) is the only possibly non-zero block of a full
\((\mathcal{S}+\mathcal{T}) \times (\mathcal{S}+\mathcal{T})\) adjacency
matrix, since links run only from source keys to target keys, never
among source keys, among target keys, or back from target to source. The
condition that the weights attached to each source key sum to one can be
checked by ensuring every row of the matrix encoding sums to one:
\(\mathbf{C}\ell_{T} = \ell_{S}\), where \(\ell_{m}\) denotes a vector
of ones of length \(m\). This row-sum property and
Proposition~\ref{prp-transform-linalg} are proved in
Section~\ref{sec-appendix-encodings}. Any row of \(\mathbf{C}\ell_{T}\)
not equal to one therefore locates a source key with at least one
incorrectly specified outgoing relation, providing additional diagnostic
detail compared to ad-hoc checks comparing the totals alone. Ad-hoc
comparison of totals for \(A\) and \(A'\) are not sufficient for
validating that the data transformations matches the intended
harmonization design because there are multiple ways to re-aggregate or
re-distribute a disaggregated mass which will preserve the total numeric
mass.

Proposition~\ref{prp-transform-linalg} recasts crossmap transforms in
terms of matrix multiplication, and provides a basis for extracting
harmonization logic from well-formed existing ex-post harmonization
scripts. For well-formed scripts, we can view the script as performing
the operation \(\mathbf{C'x}\), and extract the mapping logic by
replacing the actual input data \(\mathbf{x}\) with an identify matrix
\(\mathbf{I}\), and obtaining \(\mathbf{C'I} = \mathbf{C'}\). In
practice, it is usually not possible to replace the actual input data
\(\mathbf{x}\) directly with an identity matrix, but we can
alternatively pass though identity vectors, one for each source key,
through the script and combining the output data. The identity vector
\(o_j\) is all zeros except for the \(j\)-th key in the \(S\) element
source index set \(\mathcal{S}\). Then running the script returns a
\(T\)-length vector with the weights for any outgoing links from \(s_j\)
to elements in \(\mathcal{T}\) as shown in Section~\ref{sec-provenance}.

\emph{Conformable} crossmaps and part-to-whole array inputs can be
additionally restricted to conformable matrix dimensions, whereby the
sufficient coverage condition \(\mathcal{K} \subseteq \mathcal{S}\)
becomes an exact coverage condition \(\mathcal{K} = \mathcal{S}\), where
both must be identical ordered sets. This allows sequences of related
crossmap transform to be collapsed. For example, if we have three
related classifications, \(\alpha, \beta, \gamma\), and two crossmaps
\(\mathcal{X}_{\alpha\rightarrow\beta}\), and
\(\mathcal{X}_{\beta\rightarrow\gamma}\). The correspondence between
\(\alpha\) and \(\gamma\) can be then composed by the matrix product
\(C_{\alpha\gamma} = C_{\alpha\beta}C_{\beta\gamma}\). For larger
datasets, collapsing multiple steps eliminates intermediate computations
which reduces the time required to produce the transformed dataset.
Furthermore, collapsing sequential crossmap transforms allows for more
direct inspection of the imputed value directly in terms of the initial
source keys as shown in Figure~\ref{fig-panama}.

\subsection{Encoding Crossmaps as Edge Lists}\label{sec-edge-encoding}

The edge list representation of crossmaps shown in
Figure~\ref{fig-crossmap-transform}, and query based representation of
crossmap transforms detailed below, permits the specification,
implementation and storage of crossmap transform logic with only tabular
data structures. This tabular representation provides a basis for
extracting mapping logic from existing coding scripts, as well as
designing better ex-post harmonization workflows based on human-centered
approaches to data wrangling and analysis such as \emph{Tidy Data}
principles and the \texttt{tidyverse} suite of R packages (Wickham 2014;
Wickham et al. 2019).

Formally, the edge list encoding is \(E(s_j,t_k,w)\) with primary key
\((s_j,t_k) \in \mathcal{R}\) and attribute
\(w = w_{jk} \in \mathcal{W}\) such that each record represents a
weighted link in \(\mathcal{X}\). This representation shows how
crossmaps extend existing crosswalk approaches from the binary relation
``\(s_j\) is related to \(t_k\)'' to ternary relation ``\(s_j\)
distributes value to \(t_k\) according to \(w_{j,k}\)''. The rows in
\(E\) also correspond to non-zero entries in the matrix encoding
\(\mathbf{C}\). Furthermore, the crossmap transform detailed in
Proposition~\ref{prp-transform-linalg} can be implemented as the
database query shown in Listing~\ref{lst-sql-matvec}\footnote{Zhou and
  Ordonez (2020) show generally that properties of directed graphs can
  be obtained via matrix multiplication on the edge list encoding (see
  the paper's online supplementary material for more details).}.

\begin{codelisting}

\caption{\label{lst-sql-matvec}Query Implementation of Matrix-Vector
Multiplication. Adapted from Zhou and Ordonez (2020).}

\centering{

\begin{Shaded}
\begin{Highlighting}[]
\KeywordTok{SELECT}\NormalTok{ E.k }\KeywordTok{as}\NormalTok{ k, }\FunctionTok{sum}\NormalTok{(E.w }\OperatorTok{*}\NormalTok{ S.x) }\KeywordTok{as}\NormalTok{ y}
\KeywordTok{FROM}\NormalTok{ E }\KeywordTok{JOIN}\NormalTok{ S }\KeywordTok{ON}\NormalTok{ E.j }\OperatorTok{=}\NormalTok{ S.j}
\KeywordTok{GROUP} \KeywordTok{BY}\NormalTok{ E.k}
\end{Highlighting}
\end{Shaded}

}

\end{codelisting}%

Coverage and mass-preservation are equally native to check against the
edge list, and for the coverage condition, more so than the matrix.
Because Listing~\ref{lst-sql-matvec} implements the crossmap transform
as a join between the edge list \(E\) and the input array on the source
key, the flexibility of database joins means an incomplete crossmap does
not raise an error by default: if a source key \(s_j\) carrying value
has no matching row in \(E\), the join simply drops it, silently losing
that value from the output. Guarding against data loss only requires
checking that the source keys appearing in \(E\) are a superset of the
keys in the input array. This is a check the matrix encoding cannot
perform directly, since \(\mathcal{S}\) and \(\mathcal{T}\) are demoted
to row and column positions once encoded as \(\mathbf{C}\). By contrast,
the mass-preserving condition is directly checkable in either
representation: grouping \(E\) by source key and summing \(w\) should
return one for every \(s_j\), the edge-list equivalent of the row-sum
check in Section~\ref{sec-matrix-encoding}, subject to the same
floating-point tolerance if weights are stored numerically rather than
symbolically.

\begin{figure}

\centering{

\pandocbounded{\includegraphics[keepaspectratio]{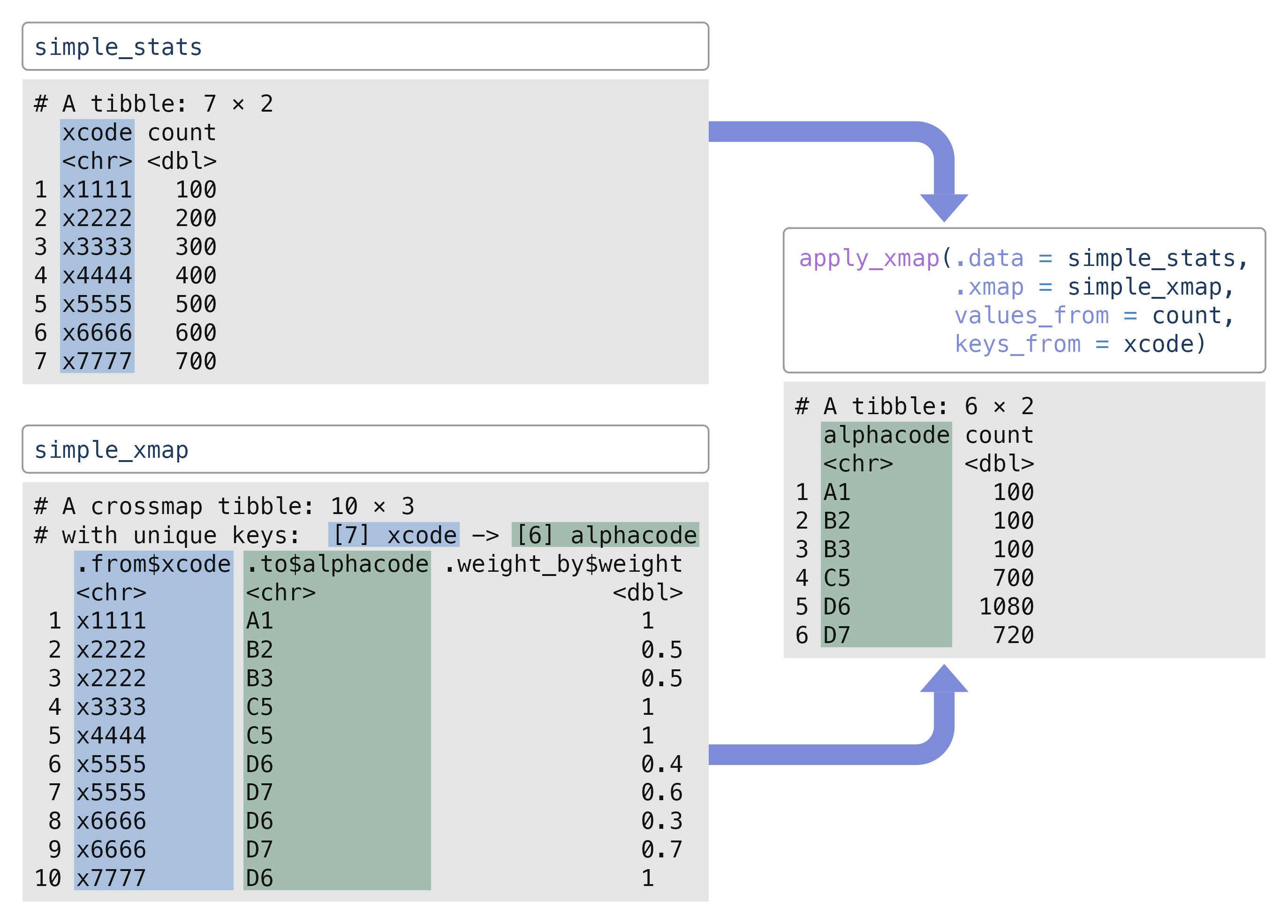}}

}

\caption{\label{fig-xmap-classes}The edge list encoding in the
\texttt{xmap} package, on the same example as
Figure~\ref{fig-crossmap-transform}. The crossmap \texttt{simple\_xmap}
and the part-to-whole array \texttt{simple\_stats} are both tables;
\texttt{apply\_xmap()} joins them on the source key and re-aggregates by
target key, returning a part-to-whole array over the target
classification.}

\end{figure}

Figure~\ref{fig-xmap-classes} shows the edge list encoding as realized
in the \texttt{xmap} package (Huang and Puzzello 2026). The crossmap is
held as a table of source key, target key and weight, the part-to-whole
array as a table of key and value, and the transform is the join of the
two. Holding both objects as tables keeps the crossmap inspectable and
editable with ordinary data manipulation tools, and means the transform
composes with the rest of a preparation script rather than sitting
outside it.

\section{Case Study: Harmonizing Industrial
Statistics}\label{sec-provenance}

We now apply the crossmaps framework to examine some of the operations
used to create the harmonized output series analyzed by Ardelean et al.
(2024). The dataset's harmonization code and quality checks\footnote{The
  code and data quality checks can be found at the online appendix:
  \url{https://l-puzzello.github.io/indstat-TPP/}.} were implemented for
harmonizing country, year observations of aggregate production
(i.e.~output/GDP) reported by sub-industries from the INDSTAT databases
published by UNIDO (United Nations Industrial Development Organization
2019). For each country and year observation in the INDSTAT 4 database,
industry level output measurements are reported directly in
International Standard Industrial Classification (ISIC) codes or in some
custom combination of them, as shown for Romania (\texttt{ROU}) in 1990
in Table~\ref{tbl-rou-1990}. Adherence or deviation from ISIC varies
both across countries and years (i.e.~the same country might use ISIC in
some years but not others). ISIC is a strictly hierarchical
classification system, whereby each code belongs to one of four nested
levels: a 1-letter section, a 2-digit division, a 3-digit group, and a
4-digit class. Data are reported at the 3- and 4-digit level (some
groups have only one class). Because ISIC is strictly hierarchical, the
reported values for a country-year form a whole and its parts: total
manufacturing output for that country-year, distributed across the
industry codes beneath it. Any redistribution across those codes must
leave the total unchanged. The original preparation code transformed
data reported under ISIC Revision 3, 3.1 or 4 into 3 digit group level
values under ISIC Revision 2, and included ad-hoc checks for missing
data and duplicate row checks, calculating reference totals to compare
the transformed series against, row count expectations and diagnostic
plots comparing the transformed data with existing reference data
prepared by Nicita and Olarreaga (2007).

\begin{table}

\caption{\label{tbl-rou-1990}Reported values for Romania, 1990. The
combined code \texttt{1531A} reports a single value for ISIC codes 1531
and 1532 jointly. Values are masked to avoid redistributing licenced
data.}

\centering{

\begin{verbatim}
# A tibble: 151 x 5
   country_name  year isiccomb  isic value
   <chr>        <dbl> <chr>    <dbl> <dbl>
 1 Romania       1990 151        151  1000
 2 Romania       1990 1511      1511  1000
 3 Romania       1990 1512      1512  1000
 4 Romania       1990 1513      1513  1000
 5 Romania       1990 1514      1514  1000
 6 Romania       1990 1520      1520  1000
 7 Romania       1990 153        153    NA
 8 Romania       1990 1531A     1531  1000
 9 Romania       1990 1531A     1532    NA
10 Romania       1990 1533      1533  1000
# i 141 more rows
\end{verbatim}

}

\end{table}%

\subsection{Describing the dataset as crossmap
transforms}\label{describing-the-dataset-as-crossmap-transforms}

The Crossmaps framework allows us to describing the harmonization
procedure and final dataset in terms of \emph{crossmap transforms},
\emph{part-to-whole arrays} and \emph{crossmaps}. Recall that the
overall goal is to have a harmonized dataset whereby each country, year
observation is expressed in a common statistical classification
standard. The full set of reported key-value pairs for Romania in 1990
shown in Table~\ref{tbl-rou-1990} form a single country, year
observation which we represent with the \emph{part-to-whole array}
object. A harmonized dataset is constructed through a collection of
crossmap transforms. Let \(\mathcal{N}\) be the country, year index for
\emph{part-to-whole array} observations in the dataset. The harmonized
dataset is the collection
\(\{A'^{n}_{[\mathcal{T},\mathbf{y}]} : n \in \mathcal{N}\}\) obtained
by applying a crossmap \(\mathcal{X}^{n}\) to each reported array
\(A^{n}_{[\mathcal{K},\mathbf{x}]}\), so there are as many crossmap
transforms as observations once those requiring no redistribution are
trivially admitted as identity crossmaps. Distinct observations may
share a crossmap, since two observations reporting the same combination
codes are harmonized by the same logic, but their arrays differ, as
these hold the reported values upon which each imputed observation is
calculated. The collection \(\{\mathcal{X}^{n}\}\) therefore describes
the imputation potentially applied across the dataset, while how much of
any harmonized value is imputed rather than observed depends on the
array to which each crossmap was applied.

\subsection{Extracting harmonization
logic}\label{extracting-harmonization-logic}

We can conduct more granular auditing of ex-post harmonization by
extracting the logic used to split values reported in combinations of
ISIC codes for a particular country. For example, as
Table~\ref{tbl-rou-1990} shows, in 1990, Romania reported the ISIC
classes \texttt{1531} and \texttt{1532} jointly under the combined code
\texttt{1531A}. Listing~\ref{lst-isiccomb-split} shows the function
\texttt{split\_isiccomb()} used in Ardelean et al. (2024) to split
combined values \texttt{isiccomb} into \texttt{isic} values.
Unfortunately, the function alone does not provide insight into which
country, year observations contained combined values, and which ones did
not. Using the matrix and graph encodings introduced in
Section~\ref{sec-matrix-encoding} and Section~\ref{sec-graph-encoding},
we can use output of this function when applied to the masked values to
extract mapping weights for each country, year. As shown in row 7 \& 8
of Table~\ref{tbl-rou-extracted}, the function
\texttt{split\_isiccomb()} split the value for \texttt{1531A} equally
between \texttt{1531} and \texttt{1532}, resulting in weights of
\texttt{0.5} for each link\footnote{The full extraction and validation
  workflow is walked through in the
  \href{https://cynthiahqy.github.io/xmap/articles/extract-validate-existing.html}{\texttt{xmap}
  package vignette ``Extracting Crossmaps from Existing Scripts''}.}.

\begin{codelisting}

\caption{\label{lst-isiccomb-split}Function to split \texttt{isiccomb}
values across their constituent \texttt{isic} codes.}

\centering{

\begin{Shaded}
\begin{Highlighting}[]
\NormalTok{split\_isiccomb }\OtherTok{\textless{}{-}} \ControlFlowTok{function}\NormalTok{(threefour\_df) \{}
  \CommentTok{\#\textquotesingle{} Helper function to split isiccomb values across isic codes}
  \CommentTok{\#\textquotesingle{} @param threefour\_df data frame with 3/4 digit values across isic \& isiccomb}

  \CommentTok{\# make list for interim tables}
\NormalTok{  interim }\OtherTok{\textless{}{-}} \FunctionTok{list}\NormalTok{()}

  \CommentTok{\# extract rows with isiccomb codes}
\NormalTok{  interim}\SpecialCharTok{$}\NormalTok{isiccomb.rows }\OtherTok{\textless{}{-}}
\NormalTok{    threefour\_df }\SpecialCharTok{\%\textgreater{}\%}
    \FunctionTok{filter}\NormalTok{(., }\FunctionTok{str\_detect}\NormalTok{(isiccomb, }\StringTok{\textquotesingle{}[:alpha:]\textquotesingle{}}\NormalTok{))}

  \CommentTok{\# test that we are not losing any data through spliting}
  \FunctionTok{test\_that}\NormalTok{(}\StringTok{"No \textasciigrave{}country,year\textasciigrave{} has more than one recorded \textasciigrave{}value\textasciigrave{} per \textasciigrave{}isiccomb\textasciigrave{} group"}\NormalTok{, \{}
\NormalTok{    rows\_w\_many\_values\_per\_isiccomb }\OtherTok{\textless{}{-}}
\NormalTok{      interim}\SpecialCharTok{$}\NormalTok{isiccomb.rows }\SpecialCharTok{\%\textgreater{}\%}
      \FunctionTok{group\_by}\NormalTok{(country, year, isiccomb) }\SpecialCharTok{\%\textgreater{}\%}
      \DocumentationTok{\#\# get  no of recorded (not NA) values for given \textasciigrave{}country, year, isiccomb\textasciigrave{}}
      \FunctionTok{summarise}\NormalTok{(}\AttributeTok{n\_obs =} \FunctionTok{sum}\NormalTok{(}\SpecialCharTok{!}\FunctionTok{is.na}\NormalTok{(value))) }\SpecialCharTok{\%\textgreater{}\%}
      \FunctionTok{filter}\NormalTok{(n\_obs }\SpecialCharTok{!=} \DecValTok{1}\NormalTok{) }\SpecialCharTok{\%\textgreater{}\%}
      \FunctionTok{nrow}\NormalTok{()}
    \FunctionTok{expect\_true}\NormalTok{(rows\_w\_many\_values\_per\_isiccomb }\SpecialCharTok{==} \DecValTok{0}\NormalTok{)}
\NormalTok{  \})}

  \CommentTok{\# calculate average value over isiccomb group for each country, year}
\NormalTok{  interim}\SpecialCharTok{$}\NormalTok{isiccomb.avg }\OtherTok{\textless{}{-}}
\NormalTok{    interim}\SpecialCharTok{$}\NormalTok{isiccomb.rows }\SpecialCharTok{\%\textgreater{}\%}
    \CommentTok{\# group isiccomb rows, replace na with 0 for averaging}
    \FunctionTok{group\_by}\NormalTok{(country, year, isiccomb) }\SpecialCharTok{\%\textgreater{}\%}
    \FunctionTok{mutate}\NormalTok{(}\AttributeTok{value =}\NormalTok{ tidyr}\SpecialCharTok{::}\FunctionTok{replace\_na}\NormalTok{(value, }\DecValTok{0}\NormalTok{)) }\SpecialCharTok{\%\textgreater{}\%}
    \CommentTok{\# split combination value over standard isic codes in isiccomb group}
    \FunctionTok{summarise}\NormalTok{(}
      \AttributeTok{avg.value =} \FunctionTok{mean}\NormalTok{(value),}
      \DocumentationTok{\#\# checking variables}
      \AttributeTok{n\_isic =} \FunctionTok{n\_distinct}\NormalTok{(isic),}
      \AttributeTok{n\_rows =} \FunctionTok{n}\NormalTok{()}
\NormalTok{    ) }\SpecialCharTok{\%\textgreater{}\%}
    \FunctionTok{mutate}\NormalTok{(}\AttributeTok{row\_check =}\NormalTok{ (n\_isic }\SpecialCharTok{==}\NormalTok{ n\_rows))}

  \CommentTok{\#  return(interim$isiccomb.avg)}

  \DocumentationTok{\#\# check n\_isic == n\_rows}
  \FunctionTok{test\_that}\NormalTok{(}\StringTok{"isiccomb split average is calculated with correct denominator"}\NormalTok{, \{}
    \FunctionTok{expect\_true}\NormalTok{(}\FunctionTok{all}\NormalTok{(interim}\SpecialCharTok{$}\NormalTok{isiccomb.avg}\SpecialCharTok{$}\NormalTok{row\_check))}
\NormalTok{  \})}

  \CommentTok{\# output processed data}
\NormalTok{  final }\OtherTok{\textless{}{-}}
    \FunctionTok{left\_join}\NormalTok{(}
\NormalTok{      threefour\_df,}
\NormalTok{      interim}\SpecialCharTok{$}\NormalTok{isiccomb.avg,}
      \AttributeTok{by =} \FunctionTok{c}\NormalTok{(}\StringTok{\textquotesingle{}country\textquotesingle{}}\NormalTok{, }\StringTok{\textquotesingle{}year\textquotesingle{}}\NormalTok{, }\StringTok{\textquotesingle{}isiccomb\textquotesingle{}}\NormalTok{)}
\NormalTok{    ) }\SpecialCharTok{\%\textgreater{}\%}
    \FunctionTok{rename}\NormalTok{(}\AttributeTok{value.nosplit =}\NormalTok{ value) }\SpecialCharTok{\%\textgreater{}\%}
    \FunctionTok{mutate}\NormalTok{(}
      \AttributeTok{value =} \FunctionTok{coalesce}\NormalTok{(avg.value, value.nosplit),}
      \AttributeTok{split.isiccomb =} \SpecialCharTok{!}\FunctionTok{is.na}\NormalTok{(avg.value)}
\NormalTok{    ) }\SpecialCharTok{\%\textgreater{}\%}
    \FunctionTok{select}\NormalTok{(country, year, isic, isiccomb, value, value.nosplit, split.isiccomb) }\CommentTok{\# not checking variables}

  \FunctionTok{return}\NormalTok{(final)}
\NormalTok{\}}
\end{Highlighting}
\end{Shaded}

}

\end{codelisting}%

\begin{Shaded}
\begin{Highlighting}[]
\NormalTok{split\_links }\OtherTok{\textless{}{-}}\NormalTok{ indstat}\SpecialCharTok{$}\NormalTok{masked\_sample }\SpecialCharTok{|\textgreater{}}
  \FunctionTok{split\_isiccomb}\NormalTok{() }\SpecialCharTok{|\textgreater{}}
  \FunctionTok{mutate}\NormalTok{(}\AttributeTok{weights =}\NormalTok{ value }\SpecialCharTok{/} \DecValTok{1000}\NormalTok{) }\SpecialCharTok{|\textgreater{}}
\NormalTok{  tidyr}\SpecialCharTok{::}\FunctionTok{drop\_na}\NormalTok{(weights)}
\end{Highlighting}
\end{Shaded}

\begin{table}

\caption{\label{tbl-rou-extracted}Extracted weights for Romania, 1990}

\centering{

\begin{verbatim}
# A tibble: 76 x 8
   country  year  isic isiccomb value value.nosplit split.isiccomb weights
   <chr>   <dbl> <dbl> <chr>    <dbl>         <dbl> <lgl>            <dbl>
 1 642      1990   151 151       1000          1000 FALSE              1  
 2 642      1990  1511 1511      1000          1000 FALSE              1  
 3 642      1990  1512 1512      1000          1000 FALSE              1  
 4 642      1990  1513 1513      1000          1000 FALSE              1  
 5 642      1990  1514 1514      1000          1000 FALSE              1  
 6 642      1990  1520 1520      1000          1000 FALSE              1  
 7 642      1990  1531 1531A      500          1000 TRUE               0.5
 8 642      1990  1532 1531A      500            NA TRUE               0.5
 9 642      1990  1533 1533      1000          1000 FALSE              1  
10 642      1990   154 154       1000          1000 FALSE              1  
# i 66 more rows
\end{verbatim}

}

\end{table}%

\begin{figure}

\centering{

\pandocbounded{\includegraphics[keepaspectratio]{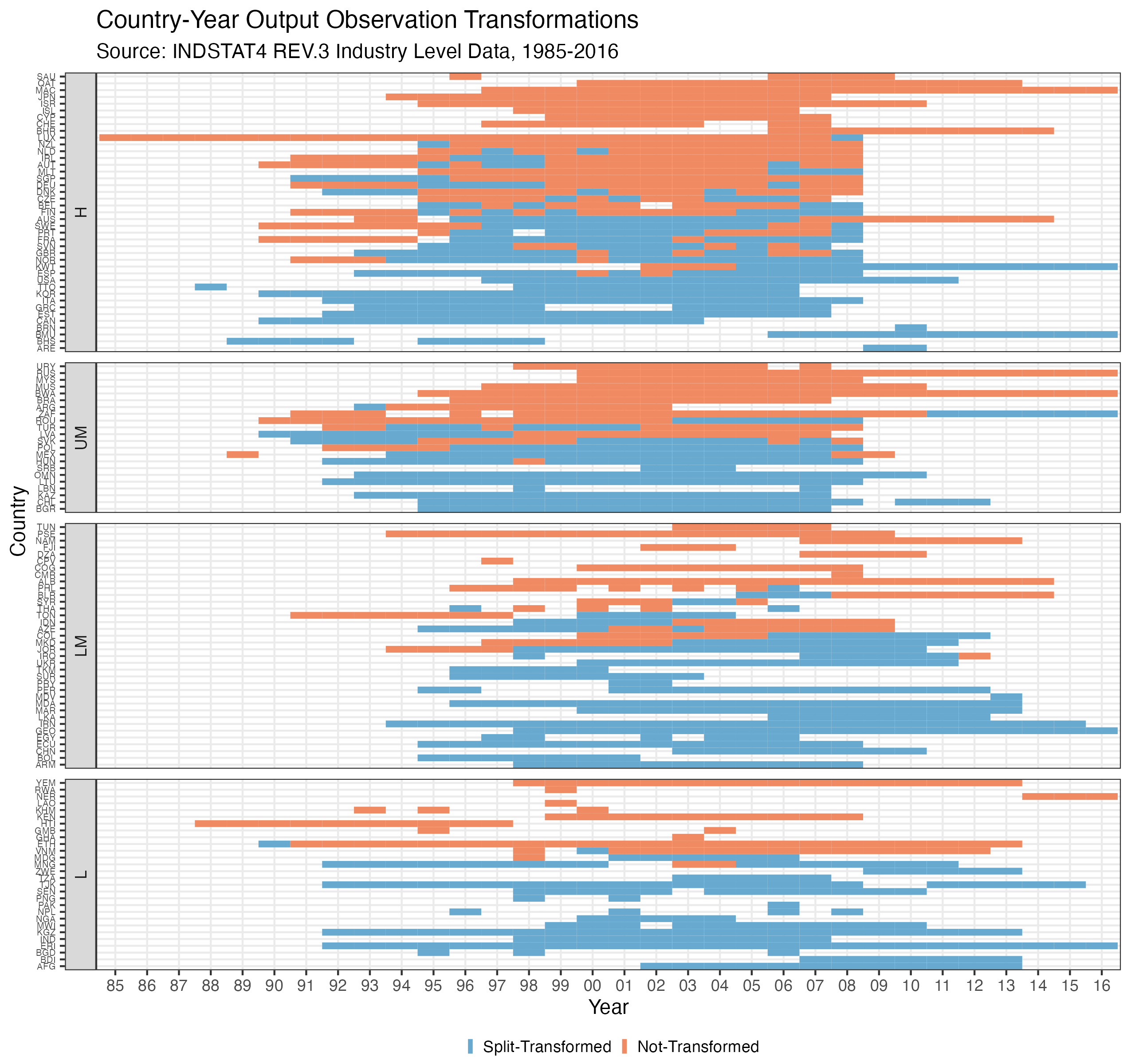}}

}

\caption{\label{fig-isiccomb-split}Summary visualization of a set of
concurrent crossmap transforms applied to industry level output
statistics collected according to country-year specific industry codes.
Each tile represents a country-year observation of output (GDP)
production in the INDSTAT4 Revision 3 Industry Level Dataset. The color
of the tile indicates whether that country-year observation contained
industry codes and associated output values that were redistributed to
the codes in the target ISIC classification}

\end{figure}

Applying the extraction to all country, year observations, we can
produce summary visualizations such as Figure~\ref{fig-isiccomb-split},
which shows which observations contained at least one value reported
using a non-standard combination of ISIC codes. We faceted the
observations by World Bank income groups (World Bank n.d.), allowing us
to see that non-standard reporting occurs across all income groups.

\subsection{Splits and aggregations}\label{splits-and-aggregations}

\begin{figure}

\begin{minipage}[t]{0.40\linewidth}

\centering{

\pandocbounded{\includegraphics[keepaspectratio]{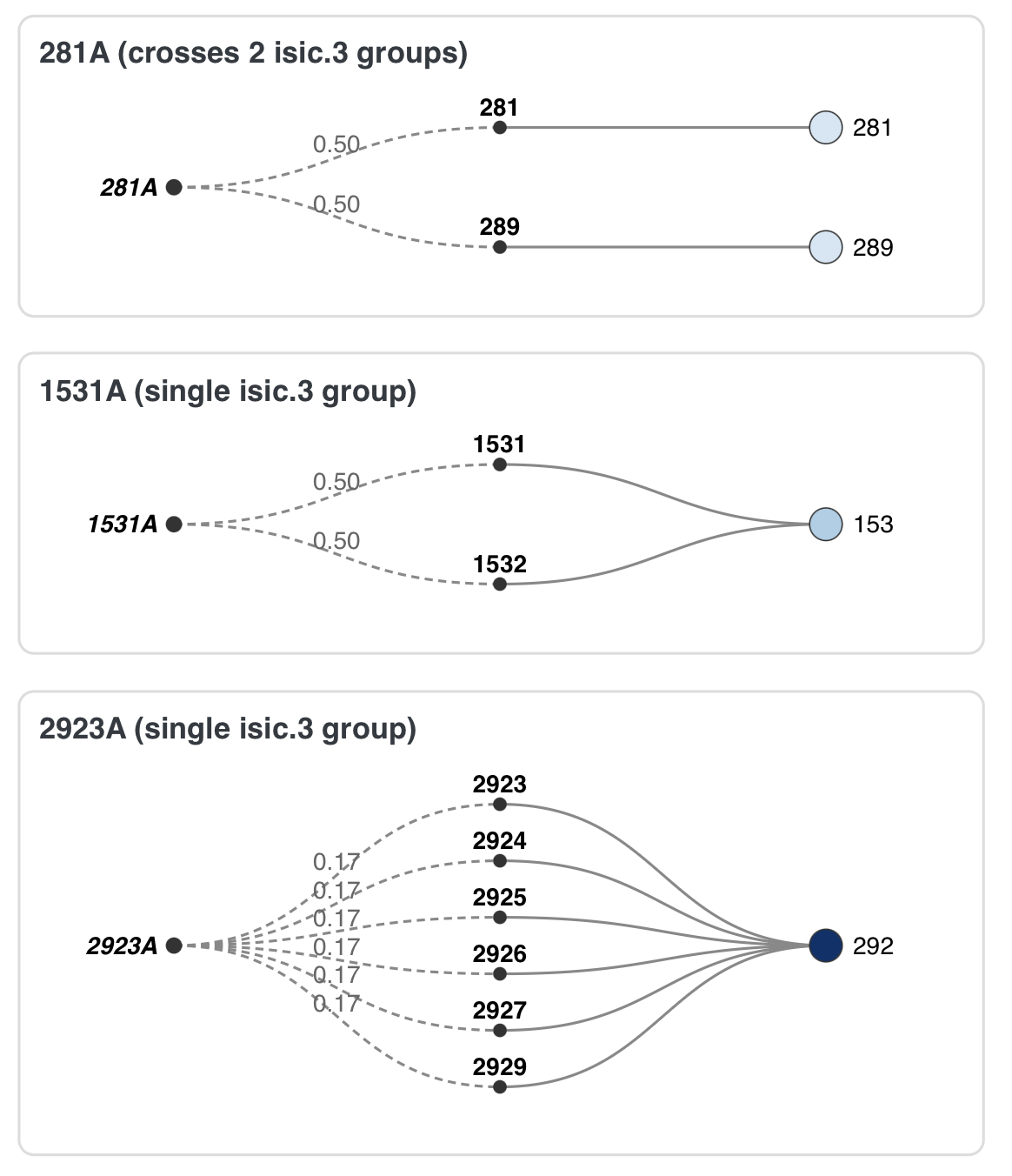}}

}

\subcaption{\label{fig-rou-seq}Sequential node-link diagram for Romania,
1990.}

\end{minipage}%
\begin{minipage}[t]{0.04\linewidth}
~\end{minipage}%
\begin{minipage}[t]{0.56\linewidth}

\centering{

\pandocbounded{\includegraphics[keepaspectratio]{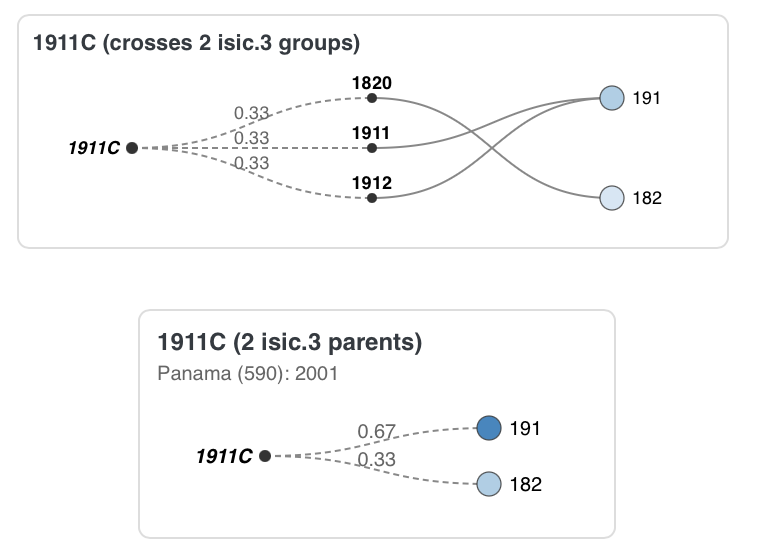}}

}

\subcaption{\label{fig-panama}Sequential (top) and collapsed (bottom)
node-link diagrams for Panama, 2001.}

\end{minipage}%

\caption[\label{fig-sequences}Node-link diagrams showing the sequence of
splitting and aggregation transforms.]{\label{fig-sequences}Node-link diagrams showing the sequence of
splitting and aggregation transforms.\footnotemark{}}

\end{figure}

Following the splitting of \texttt{isiccomb} codes, an aggregation step
from 4-digit classes to 3-digit groups was also performed. The crossmap
framework enables us to visualize and examine these two steps
sequentially or collapsed, as shown in Figure~\ref{fig-sequences}. The
converging diamond in Figure~\ref{fig-rou-seq} shows how quantifying
imputation requires tracking values through all steps of transforms, and
a case where collapsing the sequence could more accurately reflect the
actual imputation performed on a dataset. Similarly the collapsed
sequence of transformations for Panama in 2001 shown in
Figure~\ref{fig-panama} shows the effective mapping weights from
\texttt{isiccomb} to the aggregated 3 digit classes.\footnote{Examining
  and composing crossmap sequences of this kind is walked through in the
  \href{https://cynthiahqy.github.io/xmap/articles/examine-compose-crossmaps.html}{\texttt{xmap}
  package vignette: ``Examining and Composing Crossmaps''}.}

\footnotetext{The panels in
  Figure~\ref{fig-sequences} are screenshots from crossmap-explorer,
  which allows interactive inspection of the sequential and collapsed
  crossmaps for other country, year observations:
  \url{https://cynthiahqy.github.io/crossmap-explorer/03-country-variants.html}.}

\section{Limitations and Future Work}\label{sec-discussion}

\subsection{Reproducibility and reuse}\label{reproducibility-and-reuse}

The crossmap framework introduces a principled basis for standardizing,
comparing and reusing ex-post harmonization efforts. For example, rather
than only publishing crosswalks with source-target links, data
publishers and other researchers could directly share transformation
weights as crossmaps for other analysts to modify or reuse. Algorithms
for harmonization could also be restructured to output crossmaps as an
intermediate artifact, rather than only returning the final dataset.
This standard format would also allow for systematic exploration of the
analytical sensitivity and robustness of downstream analysis results to
reasonable alternative mapping weights (e.g., equal weights vs.~weights
derived from some reference data or assumptions). This systematic
variation of crossmaps could be integrated into more general frameworks
for analyzing data pre-processing decisions such as multiverse analysis
(Steegen et al. 2016), veridical data science (Yu and Kumbier 2020) or
multiphase inference (Blocker and Meng 2013).

\subsection{Extending framework
semantics}\label{extending-framework-semantics}

In Section~\ref{sec-matrix-encoding}, we suggest that missing values
should be handled as part of the source specific data preparation stage,
rather than implicitly coerced in the crossmap transform. However, there
may be cases where it is reasonable or even desirable to maintain
missing values to propagate uncertainty through to later stages of
analysis. Extending the framework to support this uncertainty
propagation requires modeling the input array \(A\) as a random
variable. The current formalization is also limited to linear
redistributions of continuous mass, but the same principles could be
extended to create data structures and formal validation conditions for
harmonization of unit-record (integer) data which require rounding.
Additionally, non-linear transformations, such as multiplicative
relations, could also be supported by generalizing the link weight to a
more general operand attribute on source-target links.

\subsection{Weight extraction, generation and
validity}\label{weight-extraction-generation-and-validity}

We show a simple example of weight extraction in
Section~\ref{sec-provenance}. However, untangling scripts with nested
conditional logic is not straightforward, and it may not always be
possible to extract intermediate transformation weights within sequence
of crossmap transforms. A possible work around may be to use large
language models to directly extract weights via some reasoning process,
or to refactor existing scripts for programmatic extraction of weights.
Large languages models can also be used to generate or propose mappings
between statistical classifications (e.g., Ma et al. 2025), for which
the crossmaps framework can be used to verify the structural validity of
generated mappings. Further work is also required to formalize the
notion of \emph{potential imputation} for an entire crossmap. This could
include properties such as the relative share of one-to-one, one-to-many
and many-to-one components, whereby crossmaps with only one-to-one
components form a zero imputation baseline from which more complex
crossmaps can be compared.

\subsection{Software usability and
development}\label{software-usability-and-development}

The prototype package \texttt{xmap} (Huang and Puzzello 2026) supports
the illustrative examples and static visualizations in this paper, but
performance at scale, usability concerns and interactivity are also not
directly addressed here. Identification of structures like the
reconvergent diamond in Figure~\ref{fig-rou-seq} in crossmap sequences
is currently only possible via visual inspection, but could be supported
programmatically. Interactive visualization could be used to support
easier comprehension and management of larger crossmaps, which can
easily contain hundreds of subgraph components, especially as every
one-to-one link forms a disjoint component. For example, less
interesting one-to-one components could be bundled, hidden or collapsed,
while components with fraction weights are left expanded for the user to
examine more closely. Such visualizations could also form the basis for
code-free interface for creating and editing crossmaps -- e.g.~by
allowing users to directly edit weights by selecting the relevant label,
or to remove links by selecting the relevant connecting lines.

\section{Conclusion}\label{sec-conclusion}

This paper contributes two declared data structures for ex-post
harmonization: the \emph{part-to-whole array} and the \emph{crossmap},
and formalized conditions for encoding and validating \emph{crossmap
transforms}, which are transformations of aggregated statistics from one
classification standard to another. We discussed and illustrated how
equivalent graph, matrix and list representations of these mappings can
reveal insights and guide novel approaches to theoretical and practical
issues in ex-post harmonization.

\section*{Acknowledgments}\label{acknowledgments}
\addcontentsline{toc}{section}{Acknowledgments}

The author thanks Laura Puzzello, Rob J. Hyndman, Sarah Goodwin, Simon
Angus, Harriet Mason, Mitchell O'Hara-Wild, Daniele Girolimetto, and
other members of Monash NUMBATs and SoDa Labs for their helpful
comments, suggestions, and contributions to this work.

\section*{Funding}\label{funding}
\addcontentsline{toc}{section}{Funding}

This work was supported by an Australian Government Research Training
Program (RTP) Scholarship, and top-up scholarships from the Monash Data
Futures Institute and the Statistical Society of Australia.

\section*{Disclosure Statement}\label{disclosure-statement}
\addcontentsline{toc}{section}{Disclosure Statement}

The author reports there are no competing interests to declare.

\section*{Data Availability
Statement}\label{data-availability-statement}
\addcontentsline{toc}{section}{Data Availability Statement}

The \texttt{xmap} package (Huang and Puzzello 2026) implements the
Crossmaps framework described in this paper and is available from CRAN
at \url{https://doi.org/10.32614/CRAN.package.xmap}, with development
sources at \url{https://github.com/cynthiahqy/xmap}. The illustrative
examples in Section~\ref{sec-framework} and Section~\ref{sec-encodings}
use data distributed with the package. Source code for the static
figures in this paper, and for the interactive tool from which the
panels in Figure~\ref{fig-sequences} are taken, is available at
\url{https://github.com/cynthiahqy/collection_crossmap-images} and
\url{https://github.com/cynthiahqy/crossmap-explorer} respectively.

The case study in Section~\ref{sec-provenance} examines harmonization
logic applied to the INDSTAT 4 database (ISIC Revision 3, 2019 edition)
published by UNIDO (United Nations Industrial Development Organization
2019). INDSTAT is licensed and the reported output values cannot be
redistributed by the author. A structural subset covering eight
countries over 1990--2013, with every reported value masked to a
constant, is distributed with the \texttt{xmap} package as the
\texttt{indstat} dataset. Because the crossmaps reported in this paper
are determined by which ISIC codes are reported in combination rather
than by the magnitudes of the reported values, all crossmaps, mapping
weights and extracted harmonization logic shown here are reproducible
from this masked subset. Readers with the appropriate licence can obtain
the unmasked source data directly from UNIDO (United Nations Industrial
Development Organization 2019). The harmonization code and data quality
checks of Ardelean et al. (2024), from which the examined logic was
extracted, are available at
\url{https://l-puzzello.github.io/indstat-TPP/}.

\section*{Generative AI Declaration}\label{generative-ai-declaration}
\addcontentsline{toc}{section}{Generative AI Declaration}

Generative AI (Claude Sonnet 5 and Claude Opus 5, Anthropic) was used in
the preparation of this manuscript for copyediting and language
refinement, including drafting bridging sentences between sections, and
for software coding assistance in developing the \texttt{xmap} package
and the code used to render Figure~\ref{fig-crossmap-transform} and
Figure~\ref{fig-xmap-classes}. All AI-assisted text and code were
reviewed and verified by the author. The author has reviewed Anthropic's
usage policies and confirms that use of these tools for the purposes
described is consistent with their terms of use and suitable for
publication. The author takes full responsibility for the integrity of
all content in this manuscript, including the accuracy of all
references.

\section{Appendix A: Proofs}\label{sec-appendix-encodings}

The framework objects and conditions used below are defined in
Section~\ref{sec-framework}.

The row-sum property stated in Section~\ref{sec-matrix-encoding} -- that
\(\mathbf{C}\ell_{T} = \ell_{S}\) for the matrix encoding \(\mathbf{C}\)
of any crossmap -- follows directly from
Definition~\ref{def-matrix-encoding} and
Definition~\ref{def-crossmap-collection}:

\begin{proof}
The result follows from the requirement that the sum of weights
originating from a given source key must total one for every source key
in a crossmap. Let
\(\mathbf{z} = \mathbf{C}\ell_{T} = [ z_{j} = \sum_{k=1}^T c_{jk}]\),
which returns the sum of each row \(j\) in \(\mathbf{C}\). By
Definition~\ref{def-matrix-encoding},
\(\sum_{k=1}^T c_{jk} = \sum_{k : (s_j,t_k) \in \mathcal{R}} w_{jk} + \sum_{k : (j,k) \notin \mathcal{R}} 0\).
By Definition~\ref{def-crossmap-collection},
\(\forall j \ \sum_{k : (s_j,t_k) \in \mathcal{R}} w_{jk} = 1\).
Therefore, \(\mathbf{z} = \ell_{S}\).
\end{proof}

Proposition~\ref{prp-transform-linalg}'s matrix-vector equivalence
follows similarly:

\begin{proof}
Let \(\mathcal{K}\) and \(\mathcal{S}\) be identical as ordered sets,
with common index \(i = 1 \dots S\). Now express
\(A_{[\mathcal{S},\mathbf{x}]}\) as a column vector
\(\mathbf{x} = [x_i]_{i = 1 \dots S}\). Then
\(\mathbf{y} = \mathbf{C}^{\top}\mathbf{x} = [\sum_{i=1}^S c_{ik} x_i]_{k = 1 \dots T}\).
By Definition~\ref{def-matrix-encoding},
\(\sum_{i=1}^S c_{ik} x_i = \sum_{i : (s_i, t_k) \in \mathcal{R}} x_i w_{ik}  + \sum_{i : (s_i, t_k) \notin \mathcal{R}} x_i 0\)
for all \(t_k \in \mathcal{T}\). Thus,
\(\mathbf{y} = [y_k = \sum_{i : (s_i, t_k) \in \mathcal{R}} w_{ik} x_i]\)
as per Definition~\ref{def-crossmap-transform}.
\end{proof}

\section*{References}\label{references}
\addcontentsline{toc}{section}{References}

\protect\phantomsection\label{refs}


\begin{CSLReferences}{1}{1}
\bibitem[\citeproctext]{ref-ardeleanGrowthVolatilityTrade2024}
Ardelean, Adina, Miguel León-Ledesma, and Laura Puzzello. 2024.
{``Growth Volatility and Trade: {Market} Diversification Vs. Production
Specialization.''} \emph{Journal of Economic Behavior \& Organization}
225 (September): 252--71.
\url{https://doi.org/10.1016/j.jebo.2024.07.001}.

\bibitem[\citeproctext]{ref-pkg-countrycode}
Arel-Bundock, Vincent, Nils Enevoldsen, and CJ Yetman. 2018.
{``Countrycode: An r Package to Convert Country Names and Country
Codes.''} \emph{Journal of Open Source Software} 3 (28): 848.
\url{https://doi.org/10.21105/joss.00848}.

\bibitem[\citeproctext]{ref-bartonicekNoMoreNo2025}
Bartonicek, Adam, Simon Urbanek, and Paul Murrell. 2025. {``No More, No
Less Than Sum of Its Parts: Groups, Monoids, and the Algebra of
Graphics, Statistics, and Interaction.''} \emph{Journal of Computational
and Graphical Statistics} 34 (3): 1063--74.
\url{https://doi.org/10.1080/10618600.2024.2429708}.

\bibitem[\citeproctext]{ref-blockerPotentialPerilsPreprocessing2013}
Blocker, Alexander W., and Xiao-Li Meng. 2013. {``The Potential and
Perils of Preprocessing: {Building} New Foundations.''} \emph{Bernoulli}
19 (4): 1176--211. \url{https://doi.org/10.3150/13-BEJSP16}.

\bibitem[\citeproctext]{ref-borsProvenanceTaskAbstraction2019}
Bors, Christian, John Wenskovitch, Michelle Dowling, et al. 2019. {``A
{Provenance Task Abstraction Framework}.''} \emph{IEEE Computer Graphics
and Applications} 39 (6): 46--60.
\url{https://doi.org/10.1109/MCG.2019.2945720}.

\bibitem[\citeproctext]{ref-cheneyProvenanceDatabasesWhy2009}
Cheney, James, Laura Chiticariu, and Wang-Chiew Tan. 2009. {``Provenance
in {Databases}: {Why}, {How}, and {Where}.''} \emph{Foundations and
Trends in Databases} 1 (4): 379--474.
\url{https://doi.org/10.1561/1900000006}.

\bibitem[\citeproctext]{ref-denkEastNeuchatelUniversal2004}
Denk, M., and K. A. Froeschl. 2004. {``East of {Neuchatel}: A Universal
Model for the Representation of Statistical Taxonomy Systems.''}
\emph{Proceedings. 16th {International Conference} on {Scientific} and
{Statistical Database Management}, 2004.} (Santorini Island, Greece),
373--82. \url{https://doi.org/10.1109/SSDM.2004.1311233}.

\bibitem[\citeproctext]{ref-dornerNovelTechnologyindustryConcordance2018}
Dorner, Matthias, and Dietmar Harhoff. 2018. {``A Novel
Technology-Industry Concordance Table Based on Linked
Inventor-Establishment Data.''} \emph{Research Policy} 47 (4): 768--81.
\url{https://doi.org/10.1016/j.respol.2018.02.005}.

\bibitem[\citeproctext]{ref-dubrowRiseCrossnationalSurvey2016}
Dubrow, Joshua Kjerulf, and Irina Tomescu-Dubrow. 2016. {``The Rise of
Cross-National Survey Data Harmonization in the Social Sciences:
Emergence of an Interdisciplinary Methodological Field.''} \emph{Quality
\& Quantity} 50 (4): 1449--67.
\url{https://doi.org/10.1007/s11135-015-0215-z}.

\bibitem[\citeproctext]{ref-ehlingHarmonisingDataOfficial2003}
Ehling, Manfred. 2003. {``Harmonising {Data} in {Official
Statistics}.''} In \emph{Advances in {Cross-National Comparison}},
edited by Jürgen H. P. Hoffmeyer-Zlotnik and Christof Wolf. Springer US.
\url{https://doi.org/10.1007/978-1-4419-9186-7_2}.

\bibitem[\citeproctext]{ref-assertr-pkg}
Fischetti, Tony. 2023. \emph{Assertr: Assertive Programming for r
Analysis Pipelines}. \url{https://docs.ropensci.org/assertr/}.

\bibitem[\citeproctext]{ref-fortierMaelstromResearchGuidelines2016}
Fortier, Isabel, Parminder Raina, Edwin R Van Den Heuvel, et al. 2016.
{``Maelstrom {Research} Guidelines for Rigorous Retrospective Data
Harmonization.''} \emph{International Journal of Epidemiology}, ahead of
print, June. \url{https://doi.org/10.1093/ije/dyw075}.

\bibitem[\citeproctext]{ref-gebruDatasheetsDatasets2021}
Gebru, Timnit, Jamie Morgenstern, Briana Vecchione, et al. 2021.
{``Datasheets for Datasets.''} \emph{Communications of the ACM} 64 (12):
86--92. \url{https://doi.org/10.1145/3458723}.

\bibitem[\citeproctext]{ref-goerlichTypologyRepresentationAlterations2018}
Goerlich, Francisco, and Francisco Ruiz. 2018. {``Typology and
{Representation} of {Alterations} in {Territorial Units}: {A
Proposal}.''} \emph{Journal of Official Statistics} 34 (1): 83--106.
\url{https://doi.org/10.1515/jos-2018-0005}.

\bibitem[\citeproctext]{ref-grandaDataHarmonization2016}
Granda, Peter, and Emily Blasczyk. 2016. {``Data {Harmonization}.''} In
\emph{Guidelines for {Best Practice} in {Cross-Cultural Surveys}}.
Survey Research Center, Institute for Social Research, University of
Michigan.

\bibitem[\citeproctext]{ref-grandaHarmonizingSurveyData2010}
Granda, Peter, Christof Wolf, and Reto Hadorn. 2010. {``Harmonizing
{Survey Data}.''} In \emph{Survey {Methods} in {Multinational},
{Multiregional}, and {Multicultural Contexts}}, 1st ed., edited by Janet
A. Harkness, Michael Braun, Brad Edwards, et al. Wiley.
\url{https://doi.org/10.1002/9780470609927.ch17}.

\bibitem[\citeproctext]{ref-huangVisualisingCategoryRecoding2023}
Huang, Cynthia A. 2023. \emph{Visualising Category Recoding and Numeric
Redistributions}. arXiv:2308.06535. arXiv.
\url{https://arxiv.org/abs/2308.06535}.

\bibitem[\citeproctext]{ref-xmap-pkg}
Huang, Cynthia A., and Laura Puzzello. 2026. \emph{{xmap}: Transforming
Data Between Statistical Classifications}.
\url{https://doi.org/10.32614/CRAN.package.xmap}.

\bibitem[\citeproctext]{ref-hulligerLinkingClassificationsLinear1998}
Hulliger, Beat. 1998. {``Linking of {Classifications} by {Linear
Mappings}.''} \emph{Journal of Official Statistics} 14 (January):
255--66.

\bibitem[\citeproctext]{ref-humlumCrosswalksDISCO88DISCO082021}
Humlum, Anders. 2021. \emph{Crosswalks {Between} ({D}){ISCO88} and
({D}){ISCO08 Occupational Codes}}. Https://www.andershumlum.com/codes.

\bibitem[\citeproctext]{ref-humlumRobotAdoptionLabor2022}
Humlum, Anders. 2022. \emph{Robot Adoption and Labor Market Dynamics}.
Rockwool Foundation Research Unit.

\bibitem[\citeproctext]{ref-pointblank-pkg}
Iannone, Richard, Mauricio Vargas, and June Choe. 2026.
\emph{Pointblank: Data Validation and Organization of Metadata for Local
and Remote Tables}. \url{https://rstudio.github.io/pointblank/}.

\bibitem[\citeproctext]{ref-kandelResearchDirectionsData2011}
{Kandel, Sean, Jeffrey Heer, Catherine Plaisant, et al.} 2011.
{``Research Directions in Data Wrangling: {Visualizations} and
Transformations for Usable and Credible Data.''} \emph{Information
Visualization} 10 (4): 271--88.
\url{https://doi.org/10.1177/1473871611415994}.

\bibitem[\citeproctext]{ref-khanMetadataCrosswalksWay2015}
Khan, Nadim Akhtar, S M Shafi, and Sabiha Zehra Rizvi. 2015. {``Metadata
{Crosswalks} as a {Way Towards Interoperability}:''} In
\emph{Encyclopedia of {Information Science} and {Technology}}, Third,
edited by Mehdi Khosrow-Pour, D.B.A. IGI Global.
\url{https://doi.org/10.4018/978-1-4666-5888-2.ch177}.

\bibitem[\citeproctext]{ref-kolczynskaMicroMacrolevelDeterminants2020}
Kołczyńska, Marta. 2020. {``Micro- and Macro-Level Determinants of
Participation in Demonstrations: {An} Analysis of Cross-National Survey
Data Harmonized Ex-Post.''} \emph{Methods, Data, Analyses} 14 (1): 36.
\url{https://doi.org/10.12758/mda.2019.07}.

\bibitem[\citeproctext]{ref-kolczynskaCombiningMultipleSurvey2022}
Kołczyńska, Marta. 2022. {``Combining Multiple Survey Sources: {A}
Reproducible Workflow and Toolbox for Survey Data Harmonization.''}
\emph{Methodological Innovations} 15 (1): 62--72.
\url{https://doi.org/10.1177/20597991221077923}.

\bibitem[\citeproctext]{ref-korenDataCodeAvailability2022}
Koren, Miklós, Marie Connolly, Joan Lull, and Lars Vilhuber. 2022.
\emph{Data and {Code Availability Standard}}. December.
\url{https://doi.org/10.5281/ZENODO.7436134}.

\bibitem[\citeproctext]{ref-concordance-pkg}
Liao, Steven, In Song Kim, Sayumi Miyano, and Feng Zhu. 2020.
\emph{Concordance: Product Concordance}.
\url{https://doi.org/10.32614/CRAN.package.concordance}.

\bibitem[\citeproctext]{ref-lucchesiSmallsetTimelinesVisual2022}
Lucchesi, Lydia R., Petra M. Kuhnert, Jenny L. Davis, and Lexing Xie.
2022. {``Smallset {Timelines}: {A Visual Representation} of {Data
Preprocessing Decisions}.''} \emph{2022 {ACM Conference} on {Fairness},
{Accountability}, and {Transparency}} (Seoul Republic of Korea), June,
1136--53. \url{https://doi.org/10.1145/3531146.3533175}.

\bibitem[\citeproctext]{ref-maCanLargeLanguage2025}
Ma, Bolei, Cynthia A. Huang, and Anna-Carolina Haensch. 2025. {``Can
{Large Language Models Advance Crosswalks}? {The Case} of {Danish
Occupation Codes}.''} \emph{Proceedings of the 2025 {Conference} of the
{Nations} of the {Americas Chapter} of the {Association} for
{Computational Linguistics}: {Human Language Technologies} ({Volume} 4:
{Student Research Workshop})} (Albuquerque, USA), 392--99.
\url{https://doi.org/10.18653/v1/2025.naacl-srw.38}.

\bibitem[\citeproctext]{ref-mackeyStrayrReadytouseAustralian2023}
Mackey, Will, Matt Johnson, David Diviny, et al. 2023. \emph{Strayr:
{Ready-to-use Australian} Common Structures and Classifications and
Tools for Working with Them}. Manual.

\bibitem[\citeproctext]{ref-nicitaTradeProductionProtection2007}
Nicita, Alessandro, and Marcelo Olarreaga. 2007. {``Trade, Production,
and Protection Database, 1976--2004.''} \emph{The World Bank Economic
Review} 21 (1): 165--71. \url{https://doi.org/10.1093/wber/lhl012}.

\bibitem[\citeproctext]{ref-niedererTACOVisualizingChanges2018}
Niederer, Christina, Holger Stitz, Reem Hourieh, Florian Grassinger,
Wolfgang Aigner, and Marc Streit. 2018. {``{TACO}: {Visualizing Changes}
in {Tables Over Time}.''} \emph{IEEE Transactions on Visualization and
Computer Graphics} 24 (1): 677--86.
\url{https://doi.org/10.1109/TVCG.2017.2745298}.

\bibitem[\citeproctext]{ref-pierceConcordanceTenDigitUS2012}
Pierce, Justin R, and Peter K Schott. 2012. {``A {Concordance Between
Ten-Digit U}.{S}. {Harmonized System Codes} and {SIC}/{NAICS Product
Classes} and {Industries}.''} \emph{Journal of Economic and Social
Measurement} 37 (1-2): 61--96.
\url{https://doi.org/10.3233/JEM-2012-03}.

\bibitem[\citeproctext]{ref-pushkarnaDataCardsPurposeful2022}
Pushkarna, Mahima, Andrew Zaldivar, and Oddur Kjartansson. 2022. {``Data
{Cards}: {Purposeful} and {Transparent Dataset Documentation} for
{Responsible AI}.''} \emph{2022 {ACM Conference} on {Fairness},
{Accountability}, and {Transparency}} (Seoul Republic of Korea), June,
1776--826. \url{https://doi.org/10.1145/3531146.3533231}.

\bibitem[\citeproctext]{ref-steegenIncreasingTransparencyMultiverse2016}
Steegen, Sara, Francis Tuerlinckx, Andrew Gelman, and Wolf Vanpaemel.
2016. {``Increasing {Transparency Through} a {Multiverse Analysis}.''}
\emph{Perspectives on Psychological Science} 11 (5): 702--12.
\url{https://doi.org/10.1177/1745691616658637}.

\bibitem[\citeproctext]{ref-unidoIndstat4Rev32019}
United Nations Industrial Development Organization. 2019.
\emph{{INDSTAT} 4: Industrial Statistics Database at the 4-Digit Level
of {ISIC} Revision 3}.
\href{https://stat.unido.org/}{Https://stat.unido.org/}; UNIDO.

\bibitem[\citeproctext]{ref-vanderlooDataValidationInfrastructure2021}
{van der Loo, Mark P. J., and Edwin de Jonge}. 2021. {``Data {Validation
Infrastructure} for {R}.''} \emph{Journal of Statistical Software} 97
(10): 1--31. \url{https://doi.org/10.18637/jss.v097.i10}.

\bibitem[\citeproctext]{ref-wangDiffLoopSupporting2022}
Wang, April Yi, Will Epperson, Robert A DeLine, and Steven M. Drucker.
2022. {``Diff in the {Loop}: {Supporting Data Comparison} in
{Exploratory Data Analysis}.''} \emph{{CHI Conference} on {Human
Factors} in {Computing Systems}} (New Orleans LA USA), April, 1--10.
\url{https://doi.org/10.1145/3491102.3502123}.

\bibitem[\citeproctext]{ref-wangNewTidyData2020}
Wang, Earo, Dianne Cook, and Rob J. Hyndman. 2020. {``A New Tidy Data
Structure to Support Exploration and Modeling of Temporal Data.''}
\emph{Journal of Computational and Graphical Statistics} 29 (3):
466--78. \url{https://doi.org/10.1080/10618600.2019.1695624}.

\bibitem[\citeproctext]{ref-wickhamReshapingDataReshape2007}
Wickham, Hadley. 2007. {``Reshaping {Data} with the Reshape
{Package}.''} \emph{Journal of Statistical Software} 21 (12): 1--20.
\url{https://doi.org/10.18637/jss.v021.i12}.

\bibitem[\citeproctext]{ref-wickhamLayeredGrammarGraphics2010}
Wickham, Hadley. 2010. {``A Layered Grammar of Graphics.''}
\emph{Journal of Computational and Graphical Statistics} 19 (1): 3--28.
\url{https://doi.org/10.1198/jcgs.2009.07098}.

\bibitem[\citeproctext]{ref-wickhamTidyData2014}
Wickham, Hadley. 2014. {``Tidy {Data}.''} \emph{Journal of Statistical
Software} 59 (10): 1--23. \url{https://doi.org/10.18637/jss.v059.i10}.

\bibitem[\citeproctext]{ref-wickhamWelcomeTidyverse2019}
Wickham, Hadley, Mara Averick, Jennifer Bryan, et al. 2019. {``Welcome
to the {Tidyverse}.''} \emph{Journal of Open Source Software} 4 (43):
1686. \url{https://doi.org/10.21105/joss.01686}.

\bibitem[\citeproctext]{ref-worldbankCountryAnalyticalHistory}
World Bank. n.d. \emph{World Bank Country and Lending Groups: Country
Analytical History}.
\href{https://datahelpdesk.worldbank.org/knowledgebase/articles/906519}{Https://datahelpdesk.worldbank.org/knowledgebase/articles/906519}.

\bibitem[\citeproctext]{ref-xiongVisualizingScriptsData2023}
Xiong, Kai, Siwei Fu, Guoming Ding, et al. 2023. {``Visualizing the
{Scripts} of {Data Wrangling} with {SOMNUS}.''} \emph{IEEE Transactions
on Visualization and Computer Graphics} 29 (6): 2950--64.
\url{https://doi.org/10.1109/TVCG.2022.3144975}.

\bibitem[\citeproctext]{ref-yuVeridicalDataScience2020}
Yu, Bin, and Karl Kumbier. 2020. {``Veridical Data Science.''}
\emph{Proceedings of the National Academy of Sciences} 117 (8):
3920--29. \url{https://doi.org/10.1073/pnas.1901326117}.

\bibitem[\citeproctext]{ref-zhouMatrixMultiplicationSQL2020}
Zhou, Xiantian, and Carlos Ordonez. 2020. {``Matrix {Multiplication}
with {SQL Queries} for {Graph Analytics}.''} \emph{2020 {IEEE
International Conference} on {Big Data} ({Big Data})} (Atlanta, GA,
USA), December, 5872--73.
\url{https://doi.org/10.1109/BigData50022.2020.9378275}.

\end{CSLReferences}
\end{document}